\documentclass[aps,pre,amsmath,amssymb,preprint,floatfix]{revtex4-1}

\usepackage{graphicx}

\def\gene/{{\rmfamily\textsc{Gene}}}
\def\kx/{\ensuremath{k_x}}
\def\krho/{\ensuremath{k_x \rho_\mathrm{s}}}
\def\t/{\ensuremath{a / c_\mathrm{s}}}
\def\npol/{\ensuremath{n_\mathrm{pol}}}

\providecommand{\e}[1]{\ensuremath{\times 10^{#1}}}

\usepackage{color}
\newcommand{\changed}[1]{{#1}}
\newcommand{\changetwo}[1]{{#1}}

\begin{document}

\title{Comparison of local and global gyrokinetic calculations of collisionless zonal flow damping in quasi-symmetric stellarators}

\author{J.~Smoniewski$^1$\footnote{smoniewski@wisc.edu}, E.~S\'anchez$^2$, I.~Calvo$^2$, M.J.~Pueschel$^{3,4,5}$, and J.N.~Talmadge$^1$}
\affiliation{$^1$University of Wisconsin-Madison, Madison, Wisconsin 53706, U.S.A.\\
$^2$Laboratorio Nacional de Fusi\'on, CIEMAT, 28040 Madrid, Spain\\
$^3$Dutch Institute for Fundamental Energy Research, 5612 AJ Eindhoven, The Netherlands\\
$^4$Eindhoven University of Technology, 5600 MB Eindhoven, The Netherlands\\
$^5$Institute for Fusion Studies, University of Texas at Austin, Austin, Texas 78712, USA}

\begin{abstract}
The linear collisionless damping of zonal flows is
calculated for quasi-symmetric stellarator equilibria
in flux-tube, flux-surface, and full-volume geometry. 
Equilibria are studied from 
the quasi-helical symmetry configuration of
the Helically Symmetric eXperiment (HSX),
a broken symmetry configuration of HSX,
and the quasi-axial symmetry geometry of
the National Compact Stellarator eXperiment (NCSX). 
Zonal flow oscillations and long-time damping
affect the zonal flow evolution,
and the zonal flow residual goes to zero for small radial wavenumber.
The oscillation frequency and damping rate depend on
the bounce-averaged radial particle drift
in accordance with theory.
While each flux tube on a flux surface is unique,
several different flux tubes in HSX or NCSX can reproduce the zonal flow damping
from a flux-surface calculation 
given an adequate parallel extent.
The flux-surface or flux-tube calculations
can accurately reproduce the full-volume long-time residual
for moderate \kx/,
but the oscillation and damping time scales are longer
in local representations,
particularly for small \kx/ approaching the system size. 
\end{abstract}

\maketitle

\section{Introduction}

Control of turbulent transport is a crucial step in the development of fusion energy. 
In many cases, zonal flows can play a role in the regulation and reduction of turbulent transport.
A zonal flow is a toroidally and poloidally symmetric $E \times B$ flow
that can be driven by electric fields that develop from fluctuations of the plasma potential,
such as is the case for most drift wave turbulence\cite{Diamond2005:PPCF-zfpr}.
The zonal flow does not drive transport itself, 
but by facilitating transfer of energy between radial wavenumbers
it can regulate the linear instability and affect turbulence saturation\cite{Makwana2012:PoP-rsmz,Makwana2014:PRL-smzt}. 
Strong zonal flows have been found to be important
in configurations such as tokamaks\cite{Lin1998:S-ttrz} or the reversed-field pinch\cite{Carmody2015:PoP-mspp,Predebon2015:PoP-itgt,Williams2017:PoP-ttzf},
and they can affect turbulence saturation in stellarators as well.\cite{Watanabe2007:NF-gszf, Faber2015:PoP-gste, Xanthopoulos2007:PRL-ngsi, Plunk2017:PRL-dtsr}

Linear zonal flow damping is often examined as a proxy for the full zonal flow evolution\cite{Xanthopoulos2011:PRL-zfdc}
and is used in models to predict turbulent transport\cite{Toda2020:JPP-rmtt,Nunami2013:PoP-rmit}. 
The Rosenbluth-Hinton model\cite{Rosenbluth1998:PRL-pfdi} provides a zonal flow residual that describes
the undamped part of the poloidal flow in a large-aspect-ratio tokamak.
This undamped flow acts to saturate drift wave turbulence, 
and the residual is used as a measure of the amplitude that zonal flows achieve
in nonlinear simulations.
In axisymmetric systems, this is commonly the case,
and the residual is sometimes used as a proxy for the resulting turbulence saturation\cite{Dimits2000:PoP-cpbt}.
However, this is unlikely to be true
if the collisionless damping to the residual is slow compared to 
the rate at which turbulence injects energy into the zonal flow. 
In non-axisymmetric devices, the radial drift of trapped particles can drive long-time damping and oscillations of the zonal flow\cite{Mishchenko2008:PoP-cdzf,Helander2011:PPCF-ozfs,Xanthopoulos2011:PRL-zfdc,Sanchez2013:PPCF-cdft,Monreal2017:PPCF-sczo},
as will be discussed in Sec.~\ref{sec:theory}.
These features can disassociate the zonal flow residual 
from saturated turbulence. 
Calculations in this paper are linear 
and do not address the transfer of energy between modes,
but can examine changes in the collisionless damping of the zonal flow.

Zonal flow damping and the driving turbulence
both depend on aspects of the magnetic geometry,
such as the rotational transform
and trapped particle regions. 
Due to the large number of parameters that can describe the plasma boundary,
the 3D shaping of stellarators offers a large parameter space to search for
configurations that can benefit 
specific turbulence or zonal flow properties.
Particularly in helical systems optimized to reduce neoclassical transport, 
stronger zonal flows may reduce turbulent transport\cite{Watanabe2008:PRL-rttz}.
However, nonlinear simulations are expensive to include in an iterative optimization loop,
and an efficient, general, linear proxy for turbulent transport would be a powerful tool.
In order to obtain such a proxy, a thorough understanding of zonal flow dynamics in stellarators is required. 

Zonal flows have been studied numerically 
\changed{in flux-tube geometry} for
the Large Helical Device\cite{Ferrando-Margalet2007:PoP-zfit} \changed{and}
\mbox{Wendelstein 7-X}\cite{Xanthopoulos2011:PRL-zfdc},
and \changed{in full-volume geometry for} 
\mbox{TJ-II}\cite{Sanchez2018:PPCF-orzf,Sanchez2013:PPCF-cdft}\changed{, 
the Large Helical Device\cite{Monreal2017:PPCF-sczo}, 
and \mbox{Wendelstein 7-X}\cite{Monreal2016:PPCF-rzft,Monreal2017:PPCF-sczo}.
As part of benchmarking gyrokinetic codes, 
full-volume linear calculations of zonal flow damping 
have been compared to local analytic theory\cite{Matsuoka2018:PoP-ntbga,Moritaka2019:P-dgpc,Monreal2016:PPCF-rzft,Monreal2017:PPCF-sczo}. 
However,} quasi-symmetric configurations \changed{are absent from previous studies,
despite the expectation that a perfectly quasi-symmetric configuration
will support an undamped zonal flow similar to a tokamak.}
A quasi-symmetric stellarator has a symmetry in the magnitude of the magnetic field $|B|$,
and the magnetic spectrum is dominated by a single mode $B_{mn}$, so that
\begin{equation}
B \approx B_{mn} \cos(n\phi-m\theta).
\end{equation}
Here, $n$ and $m$ are the toroidal and poloidal mode numbers,
and $\phi$ and $\theta$ are the toroidal and poloidal coordinates, respectively.
When the magnetic field is described by a single mode, 
the collisionless bounce-averaged drift of trapped particles from a flux surface 
goes to zero,
reducing neoclassical transport and flow damping.
Different quasi-symmetries are defined by the choice of dominant mode in the magnetic spectrum.
Quasi-polidal symmetry has a dominant $m=0$ mode,
and is not included here.
Quasi-axial symmetry has a dominant $n=0$ mode, similar to a tokamak,
as seen in Fig.~\ref{fig:ncsx_3d}. 
Quasi-helical symmetry (QHS) uses a single $n\neq0$, $m\neq0$ mode,
creating the helical shape of the $|B|$ contours in Fig.~\ref{fig:hsx_3d}.

In this paper, the zonal flow damping is numerically calculated
in flux-tube, flux-surface, and full-volume geometry representations for quasi-symmetric configurations. 
We look to understand how much geometry information is required
for an accurate determination of the zonal flow time evolution.
Although neoclassical transport and flow damping in quasi-symmetric stellarators is more similar to tokamaks than to classical stellarators,
we show that the linear zonal flow response for a realistic 
but almost quasi-symmetric geometry still resembles a classical stellarator. 

The paper is structured as follows.
Sec.~\ref{sec:theory} reviews collisionless zonal flow damping
in non-axisymmetric equilibria and introduces the
geometries and numerical tools used in this work. 
Sec.~\ref{sec:surface} identifies the differences in calculations of zonal flow damping
in full-volume, flux-surface, and flux-tube frameworks. 
In Sec.~\ref{sec:fluxtube}, calculations of zonal flow damping in flux tubes
are shown to reproduce the zonal flow residual from flux-surface calculations,
but only for sufficiently long flux tubes. 
\changed{Sec.~\ref{sec:global} presents results from 
the quasi-symmetric and broken-symmetry configurations of HSX
and compares them to the NCSX zonal flow evolution.}

\section{Collisionless zonal flow damping}
\label{sec:theory}

The Rosenbluth-Hinton model~\cite{Rosenbluth1998:PRL-pfdi} 
quantifies the long-time linear response of the zonal flow 
to a large-radial-scale potential perturbation 
in a collisionless, axisymmetric system.
The initial amplitude of the perturbation 
is reduced by plasma polarization and  
undergoes geodesic acoustic mode (GAM) oscillations 
before relaxing to a steady-state residual. 
\changed{The long-time residual zonal flow is defined 
as the ratio of the zonal potential in the long-time limit
to the initial zonal potential.} 
In a large-aspect-ratio tokamak,
the residual amplitude depends on the geometry as\cite{Rosenbluth1998:PRL-pfdi},
\begin{equation}
\frac{\varphi\changed{_{t\rightarrow\infty}}}{\varphi_0}=\frac{1}{1+1.6 q^2/\epsilon_\mathrm{t}^{1/2}}.
\label{eq:RH_residual}
\end{equation}
and can be interpreted as a measure of how strongly collisionless processes modify the zonal flow.
Here, $\varphi$ is the zonal potential, 
$\varphi_0$ is its initial amplitude, 
$q$ is the safety factor, 
and $\epsilon_\mathrm{t}$ is the inverse aspect ratio
of the flux surface of interest.
The term $1.6 q^2/\epsilon_\mathrm{t}^{1/2}$ results from
the neoclassical polarization due to toroidally trapped ions.
When the Rosenbluth-Hinton residual is high, 
collisionless zonal flow damping is small
and the system can support strong zonal flows. 
When the residual is small, damping is significant, 
and the existence of strong zonal flows will depend on strong pumping from the turbulence. 

The zonal flow response in non-axisymmetric systems is significantly modified by neoclassical effects.
The zonal flow amplitude after the polarization decay is no longer the Rosenbluth-Hinton residual. 
Instead, helical systems exhibit decay described by a timescale 
$\tau_\mathrm{c}\sim 1/|\kx/ \bar{v}_{dr}|$
to reach a residual in a long-time limit~\changed{\cite{Sugama2005:PRL-dzfh,Sugama2006:PoP-cdzf}}.
Here, \kx/ is a radial wavenumber,
and $\bar{v}_{dr}$ is the bounce-averaged radial drift velocity.  
In an unoptimized device, 
$\bar{v}_{dr}$ is large and
the zonal flow will decay quickly to a residual, 
whereas a well-optimized device will have very long decay times. 
In a perfectly symmetric device, 
no long-time decay is observed,
corresponding to the limit of infinitely long decay times.
\changed{In this case, any geometry with finite radial particle drifts
will decay to zero residual as $\kx/\rightarrow 0$. 
Defining the residual as the zonal potential at some time shorter than $t\rightarrow \infty$
necessarily involves neoclassical effects,
as discussed in Sec.~\ref{sec:global}.}
The long-time decay in helical systems could
prevent any connection between the zonal flow residual
and saturated turbulence. 
If a system takes a long time to decay, 
nonlinear energy transfer will become important before the decay has dissipated energy in the zonal mode, 
and the residual no longer relates to the zonal flow amplitude in the quasi-stationary state. 

Furthermore, an oscillation in the zonal flow 
is caused by neoclassical effects\cite{Mishchenko2008:PoP-cdzf,Helander2011:PPCF-ozfs}.
This oscillation is characterized by the radial drift of trapped particles,
as opposed to the passing particle dependence of the GAM.
Drifting trapped particles cause a radial current
that interacts with the zonal potential perturbation to cause
zonal flow oscillations.
This oscillation is damped by Landau damping on trapped particles.
The oscillation damping and frequency both increase with the radial particle drift,
or equivalently, neoclassical transport. 
For an unoptimized device, the zonal flow oscillation is of higher frequency
but damped more quickly. 
In a well-optimized device, the zonal flow oscillation is prominent due to the small damping, 
but the oscillation frequency is small compared to characteristic inverse time scales
in fully developed turbulence.
In a perfectly symmetric device, the zonal flow oscillation vanishes. 

For both stellarators and tokamaks, the zonal flow residual depends on the radial wavenumber of the zonal perturbation\changed{,
but this dependency is stronger in stellarators than in tokamaks\cite{Monreal2016:PPCF-rzft,Yamagishi2018:PPCF-nrwd,Xiao2006:PoP-swec,Jenko2000:PoP-etgd}.}
In this paper, radial wavenumbers are normalized as $k_x\rho_\mathrm{s}$,
where $\rho_\mathrm{s}$ is the ion sound gyroradius. 
The numerical calculation of the zonal flow residual in a tokamak
matches the Rosenbluth-Hinton residual as \kx/ approaches zero.
However, any geometry with finite radial particle drifts
causes the zonal flow residual to vanish as $\kx/ \rightarrow 0$,
although the long-time decay to that residual can be very slow in a well-optimized device.

Zonal flow oscillations, zonal flow damping,
and even the zonal flow residual
are further modified by the inclusion of
a radial electric field\cite{Mishchenko2012:PoP-zfsa,Sugama2009:PoP-tzfh,Sugama2010:CPP-erzf}.
The radial electric field drives coupling across field lines in the poloidal direction,
and it is likely this would be visible
in the difference between flux-tube and flux-surface calculations. 
Radial electric fields are not included here,
and their effect on calculations
in quasi-symmetric devices,
or in local and global geometry representations,
is left for future work.

\subsection{Simulations in local and global geometry representations}

A zonal flow is a toroidally and poloidally symmetric potential perturbation, 
and the local geometry anywhere on the surface
can potentially be important to determine its response.
In an axisymmetric geometry, a field line followed for one poloidal transit samples 
all unique magnetic geometry on a surface, as would 
any other field line on the same surface.
However, different field lines in a stellarator do not generally sample the same geometry.
Local geometry variations that may be important for the zonal flow
may not be sampled by a given flux tube. 
In order to investigate the representativeness of a flux tube in stellarators, 
we examine flux-surface calculations along with 
multiple flux tubes on a surface,
and extend flux tubes for multiple poloidal turns.
Extended flux tubes follow a single field line,
but are terminated after some integer number of $2\pi$ transits of the poloidal angle,
and are identified in this paper by $\npol/$ for a flux tube of $\theta=[-\npol/\pi,\npol/\pi]$.
The effect of reduced sampling by flux tubes is seen 
by comparison between different flux tubes on a surface
and to flux-surface and full-volume calculations.
The zonal flow also has a finite radial width, and 
these reduced frameworks are compared to full-volume calculations to highlight where local representations are insufficient with respect to zonal flow dynamics. 
\changed{Simulations here use a single ion species with adiabatic electrons for computational economy.
The zonal flow oscillation frequency for multi-species plasmas with kinetic electrons can be
inferred from a straightforward relation, see Ref.~\cite{Sanchez2018:PPCF-orzf}.
All time units are normalized in units of \t/,
where $a$ is the minor radius,
and $c_\mathrm{s}$ is the ion sound speed.}

\subsubsection{Full-volume geometry}
\label{sec:euterpe}

Full-volume calculations of zonal flow damping
provide the most complete representation of geometry effects
on the zonal flow.
In  this work, these calculations are carried out
with the $\delta f$ gyrokinetic particle-in-cell code \textsc{EUTERPE}\cite{Jost2001:PoP-glgs,Kornilov2004:PoP-ggts}.
\changed{The details of the zonal flow calculation are discussed 
in Refs.~\onlinecite{Monreal2017:PPCF-sczo} and \onlinecite{Monreal2016:PPCF-rzft}.}
The full-volume geometry of the fields is represented in real space
using the PEST magnetic coordinates\cite{Grimm1976:MiCPAiRaA-cmsa},
where $\phi$ is the toroidal angle, 
$s$ is the toroidal normalized flux,
and $\theta^*$ is the poloidal angle
defined such that field lines are straight.
\changed{The real rotational transform profile of each device, 
shown in Fig.~\ref{fig:iota}, 
is used in the simulation.} 
Flat density and temperature profiles are specified across the minor radius 
\changed{with $n=10^{19} \mathrm{m}^{-3}$ and $T_\mathrm{i}=T_\mathrm{e}$. 
We perform several simulations with different values of $T_\mathrm{i}=T_\mathrm{e}$  
in the range $50,100,400,1600,6400$ eV.  
For these temperatures the inverse normalized Larmor radius $a/\rho_\mathrm{s}$ 
is in the range $(63-710)$ for NCSX and $(30-169)$ for HSX configurations. 
The radial resolution of the simulations and the number of markers 
are increased as to properly resolve the zonal flow structure 
while keeping the ratio of modes to number of markers constant.}
The plasma potential is computed from the charge density of particles
in a set of flux surfaces using B-splines. 
The potential is Fourier-transformed at each flux surface
and can be filtered in Fourier space. 
From the Fourier spectrum, \changed{only} the \changed{$(0,0)$} component 
\changed{is of interest for zonal flow calculations}
and \changed{is} extracted at individual flux surfaces.

The linear properties of the zonal flow
are extracted from the time trace of the zonal component.
These linear zonal flow relaxation simulations 
are initialized with a flux-surface-symmetric perturbation to the ion distribution function.
The initial condition has a Maxwellian velocity distribution and a radial structure such that 
a perturbation to the potential containing a single radial mode, 
$\phi \propto \cos(k_s s)$,
is produced after solving the quasi-neutrality equation.
The simulation is linearly and collisionlessly evolved, 
and the time evolution of the zonal potential at fixed radial positions is recorded. 
\changed{A long-wavelength approximation valid for $\krho/ < 1$ 
is used to simplify the quasi-neutrality equation.
The function $\Gamma_0 ( \mathrm{x} )=e^{- \mathrm{x}} I_0 ( \mathrm{x} )$ is approximated as $\Gamma_0(\mathrm{x}) \sim 1 - \mathrm{x}$, 
where $\mathrm{x} = k_x^2 \rho_\mathrm{s}^2$ and $I_0$ 
is the modified Bessel function\cite{Monreal2017:PPCF-sczo}. 
Then the quasi-neutrality equation for adiabatic electrons reads:
\begin{equation}
q_\mathrm{i} \langle n_\mathrm{i} \rangle  - \frac{e n_0 (\varphi - \{\varphi\}_s)}{T_\mathrm{e}}= -\nabla \frac{m_\mathrm{i} n_0}{B^2} \nabla_{\perp} \varphi
\end{equation}
where $n_0$ is the equilibrium density, $B$ is the magnetic field, $\langle n_i \rangle$ is the gyroaveraged ion density, $T_\mathrm{e}$ is the electron temperature, and $e$ and $m_\mathrm{i}$ are the electron charge and the ion mass, respectively. The $\{\}_s$ brackets represent a flux surface average.}

Linear zonal flow relaxation simulations
are less numerically intensive than turbulence simulations\cite{Sanchez2020:JPP-ngps}\changed{,
where many modes are allowed to evolve an interact in a nonlinear simulation,
and time steps are shorter to account for fast phenomena. 
Fewer Fourier modes are required for zonal flow calculations,
where there are no temperature and density gradients to drive turbulence 
and only a single mode is examined,
as opposed to the mode spectrum of a nonlinear simulation.
For a zonal flow calculation, a larger number of modes would only increase the numerical noise and require more computational resources.}
However, a long simulation time is required 
to extract the long-time properties of the zonal flow evolution,
which prohibits the use of full-volume calculations in an optimization loop. 
Simulations presented in this work are carried out 
retaining 6 poloidal and toroidal modes, 
with radial resolutions ranging from 24 to 64 points
\changed{to account for the radial structure of the mode}, 
and from 50\e{6} to 200\e{6} markers. 
\changed{These resolutions are similar to those in 
previous \textsc{EUTERPE} zonal flow calculations\cite{Monreal2016:PPCF-rzft,Monreal2017:PPCF-sczo}.}

\subsubsection{Flux-tube and flux-surface geometry}
\label{sec:gene_calcs}

The gyrokinetic $\delta f$ continuum code \gene/\cite{Jenko2000:PoP-etgd}
is used for calculations of the zonal flow decay
in flux-tube and flux-surface representations,
constructed from VMEC equilibria with the GIST code\cite{Xanthopoulos2006:PoP-ccng}.
All flux-tube and flux-surface calculations in this work use the $s=0.5$ flux surface,
and are compared to results from full-volume calculations about this surface.
The flux tube is a reduced-geometry representation for toroidal magnetic geometries,
and is constructed from a sheared box around a single field line
identified by a field line label in PEST\cite{Grimm1976:MiCPAiRaA-cmsa} coordinates as $\alpha = \left( \sqrt{s_0} / q_0 \right) \left( q \theta^* - \phi \right)$.
When the flux tube is centered on the outboard midplane,
$\alpha$ is also a toroidal coordinate of the center point of the flux tube. 
The box uses non-orthogonal coordinates $x$ in the radial direction, 
$y$ in the flux surface,
and $z$ along the field line.
In a \gene/ flux tube, 
a spectral representation is used for the $x$ and $y$ directions,
while the $z$ direction is in real space.
\changed{For $k_y = 0$ zonal modes,
boundary conditions in $x$, $y$, and $z$ are periodic.}
\changed{In axisymmetric systems, a flux tube of one poloidal turn 
samples all unique geometry on the flux surface.
In a non-axisymmetric system, \changetwo{a flux tube
does not necessarily close upon itself. 
The standard approach to using flux tubes in stellarators 
does not require true geometric periodicity of the flux tube\cite{Martin2018:PPCF-pbct}.} 
A stellarator-symmetric flux tube, or one that is symmetric about the midpoint $z=0$,
provides continuous, but not necessarily smooth, geometry at the flux tube endpoints.
\changetwo{However, $k_z=0$ modes, such as zonal flows, 
may be sensitive to the geometry at this boundary.
True geometric periodicity requires 
that $q \npol/ N$ is an integer,
where $N$ is the toroidal periodicity of the geometry.} 
We treat the flux tube length \npol/ as a parameter subject to convergence\changetwo{,
and show in Sec.~\ref{sec:fluxtube} that the choice of a truly periodic 
or a standard stellarator symmetric flux tube does not affect the outcome 
of the studies conducted in this work.}
} 

A flux-surface representation 
\changed{discretizes the $y$ direction in real space instead of Fourier space. 
The $z$-direction is aligned to the magnetic field,
and field lines are followed for one poloidal turn.
Calculations here use 64 $y$ points equally spaced in $\alpha$.}
A flux-tube calculation only includes the local magnetic geometry coefficients along the field line,
while a flux-surface calculation captures the variation in geometry with $\alpha$.
\changed{The radial computational domain is set by 
the magnetic shear of the configuration. 
The configurations considered in this paper have a small magnetic shear,
setting the minimum radial wavenumber for flux-surface calculations to
$k_{x,\mathrm{min}}=0.009$ in HSX and
$k_{x,\mathrm{min}}=0.158$ in NCSX.
Calculations at an appropriate $k_x$ to compare to flux-tube and full-volume calculations 
proved unfeasible in HSX due to the very small $k_{x,\mathrm{min}}$. 
Therefore, full-surface calculations are only presented in NCSX.}

The zonal flow damping, and resulting residual and oscillations, is calculated
by initializing a flux-surface-symmetric impulse to the distribution function
at a single radial mode
and allowing the amplitude to collisionlessly decay 
due to classical and neoclassical polarization
without further energy input.
In \gene/, the zonal perturbation is implemented by initializing
only one $k_x\neq 0$, $k_y=0$ mode.
\changed{In a Fourier representation, the wavenumber is explicit,
and a long-wavelength approximation is not used.}
The perturbation is introduced in the density with Maxwellian velocity space, 
which produces an equivalent potential perturbation to that used in \textsc{EUTERPE}.
Note that for the present case of adiabatic electrons, 
the two initial conditions discussed in Ref.~\cite{Pueschel2013:PoP-phmn} are identical.

\begin{figure*}
  \includegraphics[width=\linewidth]{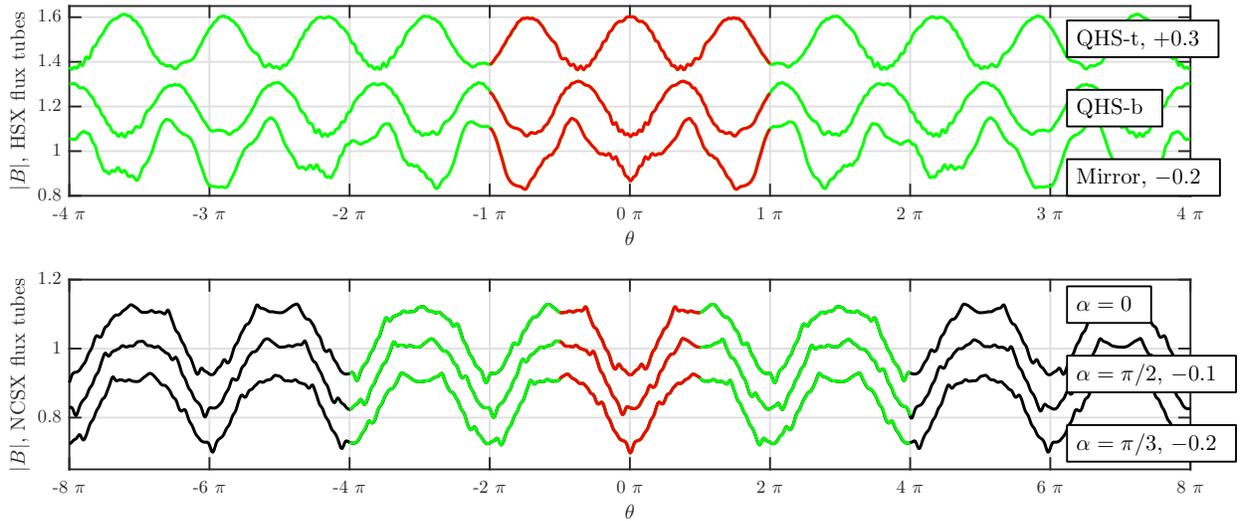}
  \changed{\caption{Comparison of $|B|$ in the various flux tubes from HSX and NCSX. A flux tube of length $\npol/=1$ is plotted in red, $\npol/=4$ in green, and $\npol/=8$ in black. Curves are shifted to avoid overlap.  \label{fig:modb}}}
\end{figure*}

Numerical calculations for linearly stable systems without dissipation
may have to contend with numerical recurrence phenomena.
Such recurrence, which results from a reestablishment of phases from the initial condition
and concomitant unphysical temporary increase in amplitudes,
can be eliminated by including numerical spatial or velocity hyper-diffusion \cite{Pueschel2010:CPC-rndg}.
However, numerical diffusion is not an appropriate solution
in nearly quasi-symmetric stellarators,
as calculation times are very long
and even a small amount of diffusion will cause significant damping of the zonal-flow residual.
To solve the problem, the parallel velocity space grid spacing $\Delta v_\parallel$
can be decreased sufficiently that the recurrence time,
\changed{$\tau_\mathrm{rec} = 2 \pi / ( k_z \Delta v_\parallel )$},
exceeds the duration of the simulation\cite{Dannert2004:CPC-vsks}.
In the present work, most flux-tube calculations
\changed{use $N_{v\parallel} > 256$ to discretize the velocity space 
spanning $v_\parallel=[-3v_{Ti},-3v_{Ti}]$,
leading to $\Delta v_\parallel = 0.015 \; v_{Ti}$.
Here, $v_{Ti} = (2 T_\mathrm{i} / m_\mathrm{i})^{1/2} $ is the ion thermal velocity.
We take $k_z$ to be the wavenumber of the periodicity of the magnetic structure,
as seen in Fig~\ref{fig:modb},
which leads to $k_z \approx 0.4 \; a^{-1}$ in HSX 
and $k_z = 0.34 \; a^{-1}$ in NCSX.
For $N_{v\parallel} = 256$, $\tau_\mathrm{rec} \approx 670 \; \t/$ in HSX
and $\tau_\mathrm{rec} \approx 790 \; \t/$ in NCSX.
This effect is seen in Fig~\ref{fig:cmp_QA_convergence}.
For $N_{v\parallel} = 384$, $\tau_\mathrm{rec} > 1000 \; \t/$ in both configurations.}

\subsection{The HSX and NCSX geometries}
\label{sec:geometry}

The zonal flow response is studied by means of 
flux-tube, flux-surface, and full-volume gyrokinetic simulations in
the Helically Symmetric eXperiment (HSX)\cite{Anderson1995:FT-hseg}
and the National Compact Stellarator eXperiment (NCSX)\cite{Neilson2002:-pdnc}.
VMEC\cite{Hirshman1983:TPoF-smmt} is used to calculate the HSX and NCSX equilibria.

\begin{figure}
  \includegraphics[width=\linewidth]{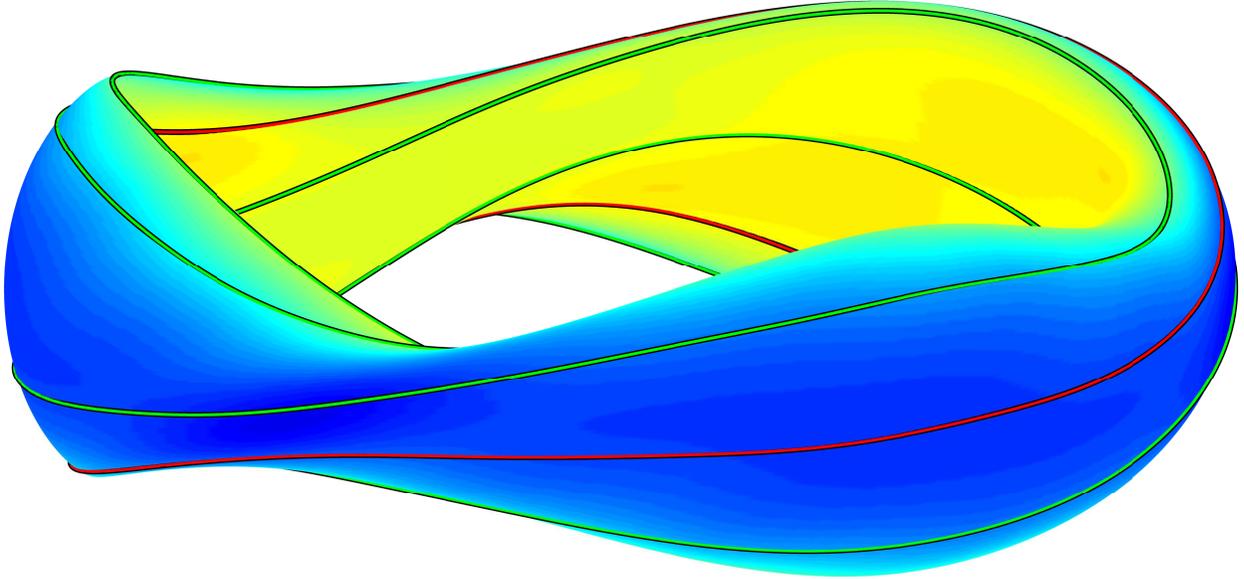}
  \caption{A flux surface and $\alpha=0$ flux tube for the $s=0.5$ surface of the NCSX configuration. Colors correspond to $|B|$, where blue is the minimum field strength. A flux tube of one poloidal turn is shown in red, and one of 4 poloidal turns in green. \label{fig:ncsx_3d}}
\end{figure}

\begin{figure}
  \includegraphics[width=\linewidth]{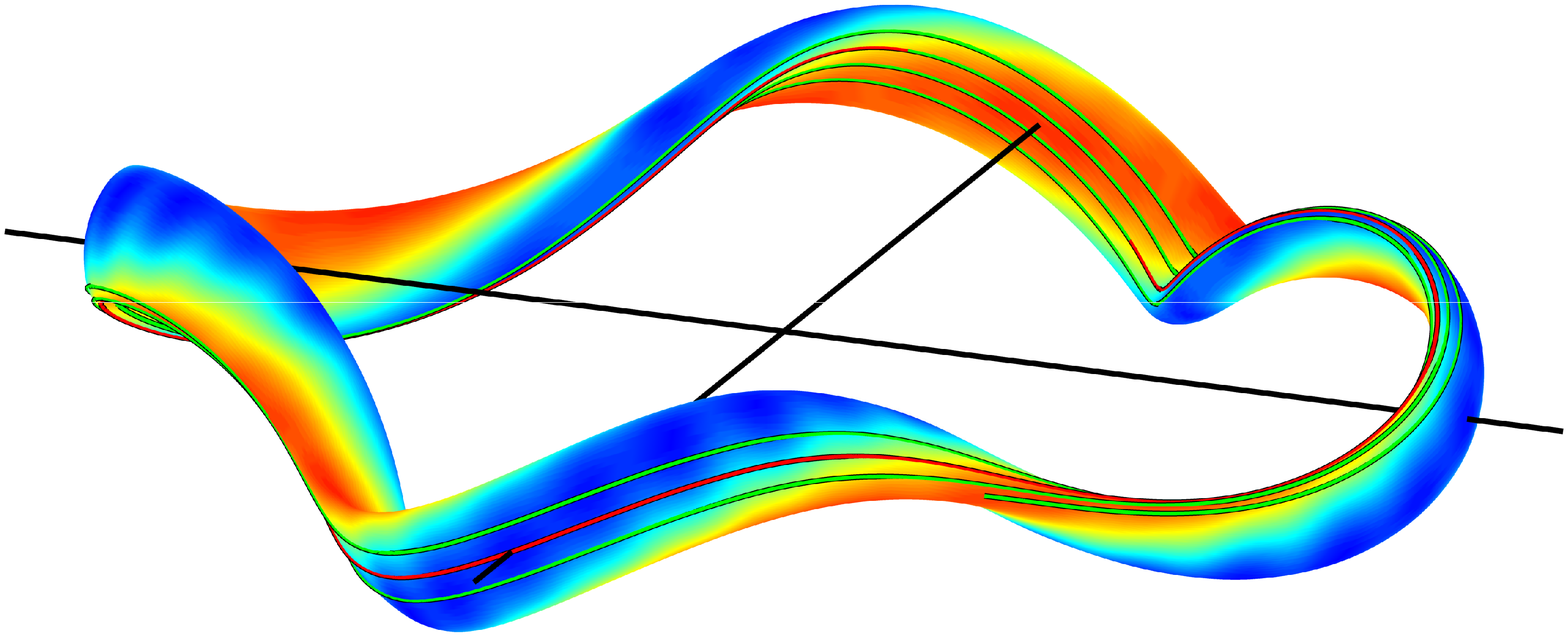}
  \caption{A flux surface and $\alpha=0$ flux tube for the $s=0.5$ surface of the QHS configuration of HSX. Colors correspond to $|B|$, where blue is the minimum field strength. A flux tube of one poloidal turn is shown in red, and one of 4 poloidal turns in green. \label{fig:hsx_3d}}
\end{figure}

\begin{figure}
  \includegraphics[width=\linewidth]{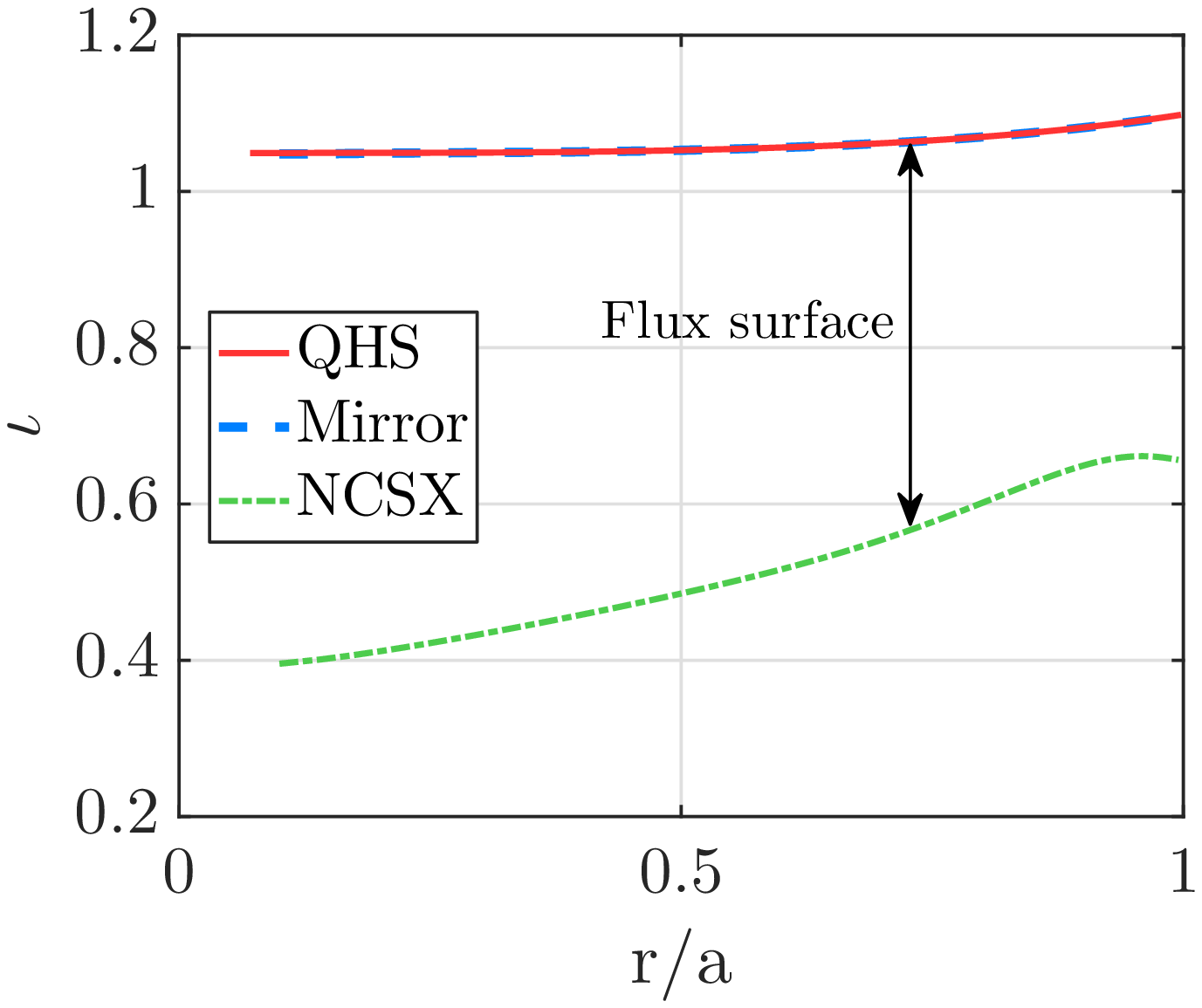}
  \changed{\caption{The iota profile in the three configurations studied. The radial location of flux-tube and flux-surface calculations is marked with the arrow. The iota profile in the HSX QHS and Mirror configurations shows a negligible difference. \label{fig:iota}}}
\end{figure}

HSX is a four-field-period optimized stellarator,
designed to improve single-particle confinement through quasi-helical symmetry.
The main coils produce the helically symmetric field,
and a set of auxiliary coils can be energized to modify the magnetic spectrum. 
The symmetry is broken by adding mirror terms to the spectrum, 
in which case transport is similar to a classical stellarator with large neoclassical transport in the low collisionality regime. 
HSX has demonstrated reduced neoclassical flow damping\cite{Gerhardt2005:PRL-eerp} and transport\cite{Canik2007:PRL-edin} in the QHS configuration.
It has also been hypothesized that neoclassical optimization may reduce anomalous transport through stronger zonal flows\cite{Watanabe2008:PRL-rttz}.
Zonal flows are clearly observed in nonlinear gyrokinetic simulations with \gene/,
but measurements are the subject of ongoing research\cite{Wilcox2011:NF-mblc}.

In configurations examined here,
equilibria are constructed from vacuum fields,
which is consistent with the very low plasma pressure in HSX. 
There are two unique flux tubes centered on the outboard midplane
that are symmetric about the midpoint $z=0$. 
\changed{The QHS-b ``bean''} flux tube ($\alpha = 0$) \changed{is} centered on the outboard midplane
of the bean\changed{-shaped} cross section,
\changed{where it} is the low-field \changed{and bad-curvature} side.
The \changed{QHS-t ``triangle''} flux tube ($\alpha = \pi/N$, with $N=4$ for HSX)
is centered on the outboard midplane in the triangle cross section,
\changed{where it} is the high-field \changed{and good-curvature} side in HSX.

The NCSX configuration was also optimized for neoclassical transport,
but is a three-period device designed for quasi-axial symmetry.
The equilibrium used here has total normalized plasma pressure $\beta \approx 4\%$.
In the NCSX configuration, we use the flux tubes symmetric about the midpoint $z=0$
($\alpha = 0$ and $\alpha = \pi/N$, with $N=3$ for NCSX),
as well as one non-symmetric flux tube ($\alpha = \pi/2$). 
The radial particle drift 
for the surface at $s=0.5$
is between that of the QHS and Mirror configurations of HSX. 
\changed{The rotational transform in NCSX is roughly half that in HSX, 
as seen in Fig.~\ref{fig:iota}}.
The difference in rotational transform
means that the part of the surface sampled by a flux tube is very different.
\changed{In Fig.~\ref{fig:hsx_3d}, the multiple turns of a flux tube in HSX
cluster together in a band around the device. 
In Fig.~\ref{fig:ncsx_3d}, the multiple turns of a flux tube in NCSX 
spread out across the surface,
potentially sampling larger variation in a shorter flux-tube length.
However, the flux tube does not allow poloidal communication between turns,
and poloidally neighboring geometry can only affect the zonal flow damping 
in a flux-tube calculation through parallel physics.}

\subsection{Fitting zonal flow oscillations and residuals}

The zonal flow decay in a stellarator includes additional
long-time damping and zonal flow oscillations
as compared to the tokamak case.
Previous studies have commonly focused on the zonal flow residual or oscillation frequency,
but there is also the short-time damping due to the polarization drift,
additional long-time damping due to the polarization of trapped particles,
the GAM oscillation,
and the zonal flow oscillation amplitude and damping.
Following Monreal\cite{Monreal2017:PPCF-sczo},
curve fitting is used to extract 
the residual and parameters of the zonal flow oscillation.
The time evolution during the post-GAM phase is described by,

\begin{equation}\label{eq:zf_eq}
\frac{\varphi_k(t)}{\varphi_k(0)} = A_\mathrm{ZF}\cos(\Omega_\mathrm{ZF}t)\mathrm{e}^{-\gamma_\mathrm{ZF}t} + R_\mathrm{ZF} + C\mathrm{e}^{-\gamma}.
\end{equation}
The zonal flow oscillation is parameterized by 
an amplitude $A_\mathrm{ZF}$,
oscillation frequency $\Omega_\mathrm{ZF}$,
and damping rate $\gamma_\mathrm{ZF}$.
The long-time decay follows
an algebraic decay
according to Ref.~\cite{Helander2011:PPCF-ozfs},
but is well approximated by an exponential decay $C\mathrm{e}^{-\gamma}$
to avoid an abundance of fitting parameters. 
The zonal flow residual is $R_\mathrm{ZF}$.
The evolution of normalized zonal potential and normalized zonal electric field are equivalent
for a zonal potential with only a single \kx/\cite{Monreal2017:PPCF-sczo}. 
However, in practice, the zonal electric field is preferred in global \textsc{EUTERPE} simulations
to simplify the fitting.

We are primarily interested in the zonal flow oscillation here,
not the GAM. 
The damping of the GAM increases with decreasing rotational transform\cite{Sugama2006:PoP-cdzf}.
In HSX, the rotational transform is about one
and the GAM is damped on time scales of the order 10 \t/.
GAM oscillations are more apparent in NCSX calculations,
as the rotational transform is about twice that in HSX,
but are still damped quickly compared to the zonal flow oscillations. 
Fitting starts after the GAM oscillations have damped away
to avoid further complexity in curve fitting.
The zonal flow oscillation is Landau-damped by trapped particles,
and depends on the radial drift off of the flux surface\cite{Helander2011:PPCF-ozfs}.
Neoclassically optimized devices can have long-lived zonal flow oscillations
as neoclassical transport is reduced,
which also reduces the oscillation frequency to well below the GAM frequency.
In the configurations examined here,
fitting the zonal flow oscillations is important to accurately fit the zonal flow residual.

\section{Comparison of local and global calculations}
\label{sec:surface}

\begin{figure}
  \includegraphics[width=\linewidth]{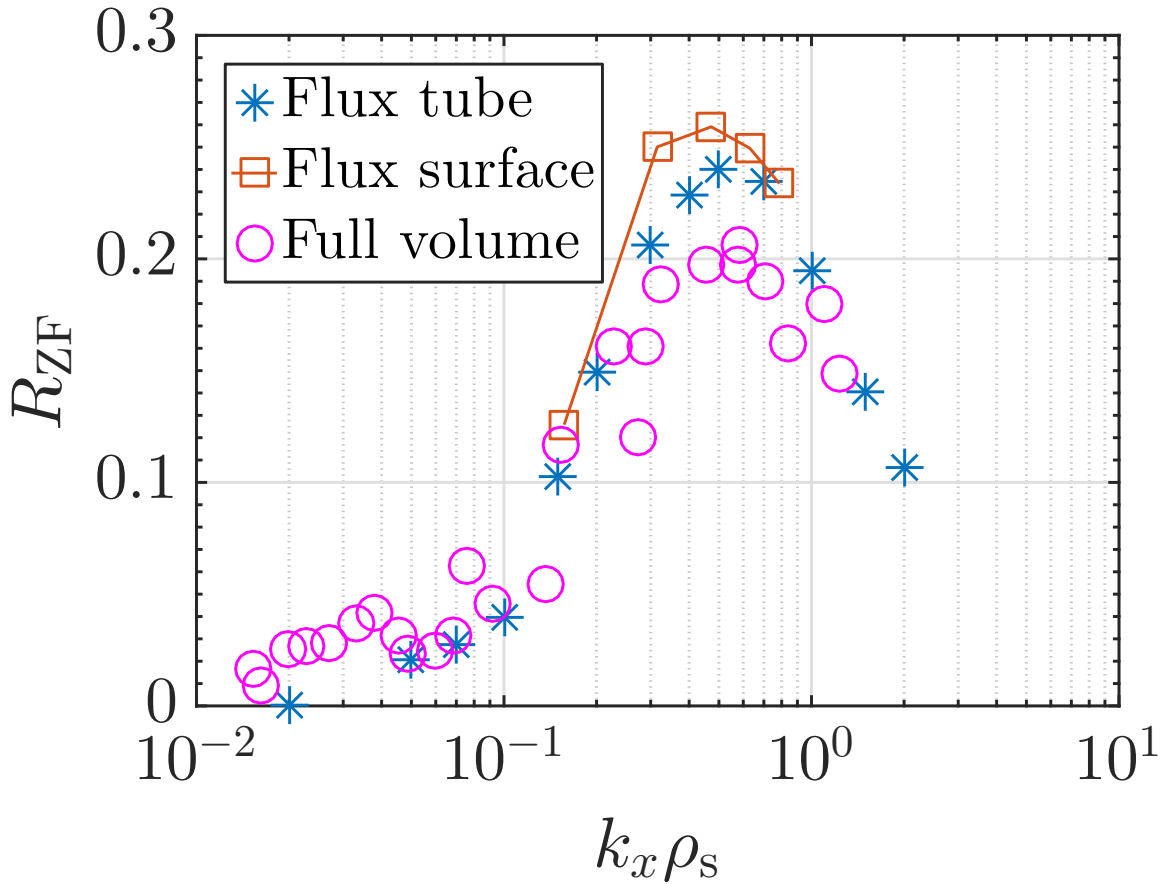}
  \caption{The zonal flow residual in the NCSX configuration from flux-tube, flux-surface, and full-volume calculations. Local and global representations largely show good agreement on the \kx/ dependence of the residual, with moderate deviations observed at very small \krho/ and near the peak residual, as expected from model limitations.\label{fig:QA_Rzf}}%
\end{figure}

The NCSX configuration is quasi-axisymmetric,
which, among three-dimensional geometries,
most closely resembles a tokamak.
As discussed in Sec.~\ref{sec:theory},
the zonal flow residual as $\kx/ \rightarrow 0$
is a key difference between symmetric and non-symmetric systems.
The time traces for full-volume, flux-surface, and flux-tube simulations
are fitted to extract the zonal flow residual plotted in Fig.~\ref{fig:QA_Rzf}.
A discussion of different flux tubes in the NCSX configuration is provided in Sec.~\ref{sec:fluxtube_ncsx}.
In NCSX, the zonal flow residual goes to zero for small \kx/,
just as it does for classical stellarators. 
The limit as $\kx/\rightarrow 0$,
as well as a peak residual around $\krho/\approx 0.5$,
is reproduced in all three geometry representations. 
However the amplitude of the residual differs between
the local and global representations particularly for very small \kx/.
In the full-volume calculations,
the peak of the residual is slightly lower,
while the smallest \kx/ support a larger residual 
than the flux-tube calculations.
A long-wavelength approximation valid for $\krho/ < 1$ is used in the global simulations,
which may be approaching its limit of validity towards the peak.
\changed{Without flux surface calculations at small \kx/,
we cannot constrain the physical cause for differences
between full-volume and flux-surface results. 
C}oupling between surfaces may be occurring,
\changed{but the same disagreement is not seen for HSX configurations in Fig.~\ref{fig:cmp_qhs_f14}.}
More importantly, the flux-surface approximation will break down when scales
are large enough to involve profile effects,
and the smallest \kx/ values examined approach the machine size. 
Thus the observed discrepancies are to be expected 
given the limitations of the frameworks. 

\begin{figure}
  \includegraphics[width=\linewidth]{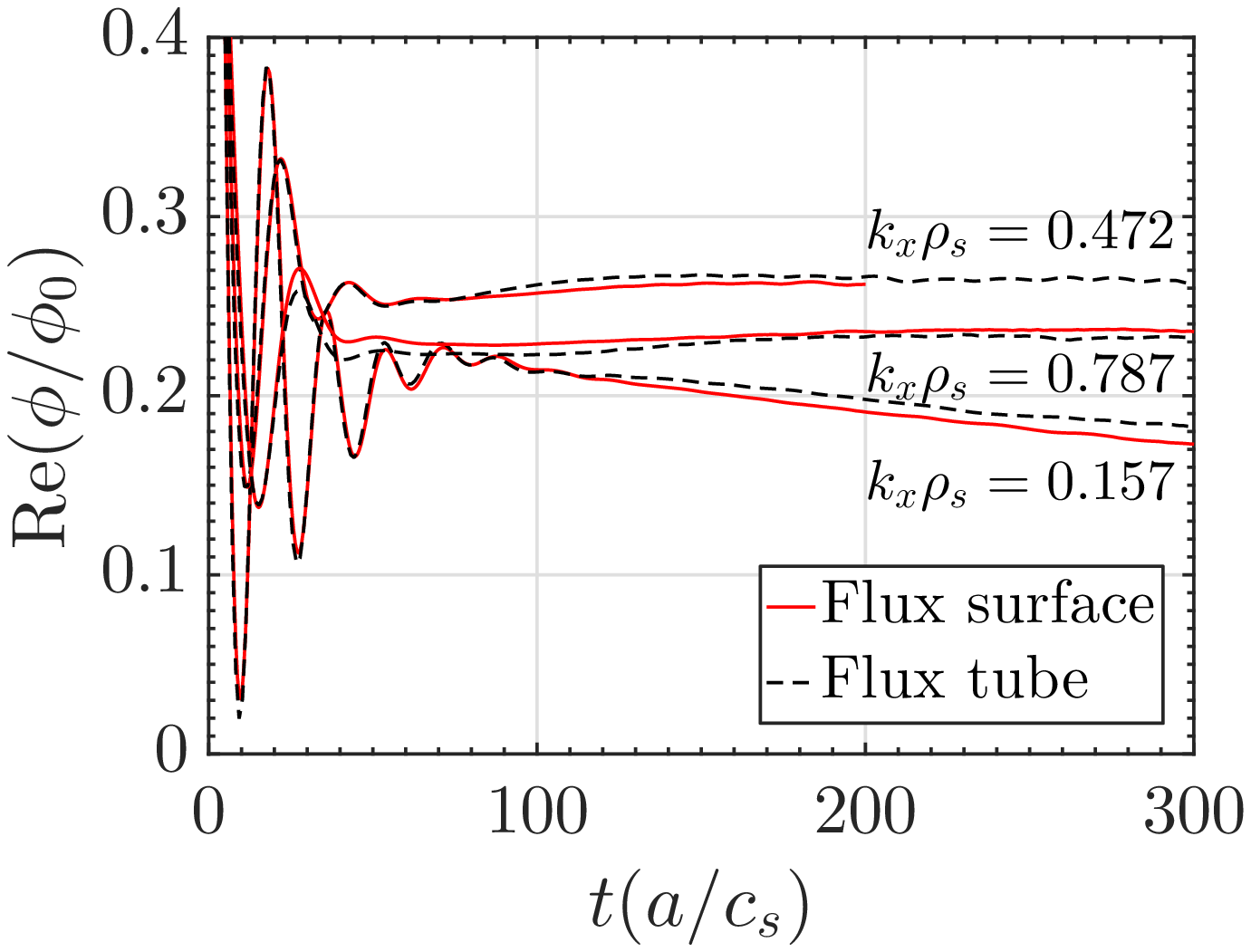}
  \caption{Comparison of zonal flow evolution in NCSX for flux-surface and flux-tube calculations. Good agreement is found for the initial polarization drift, the GAM oscillations and damping, and the long-time decay. Flux-tube calculations here use $\npol/=4$.\label{fig:qa_surface_t}}
\end{figure}

\begin{figure}
  \includegraphics[width=\linewidth]{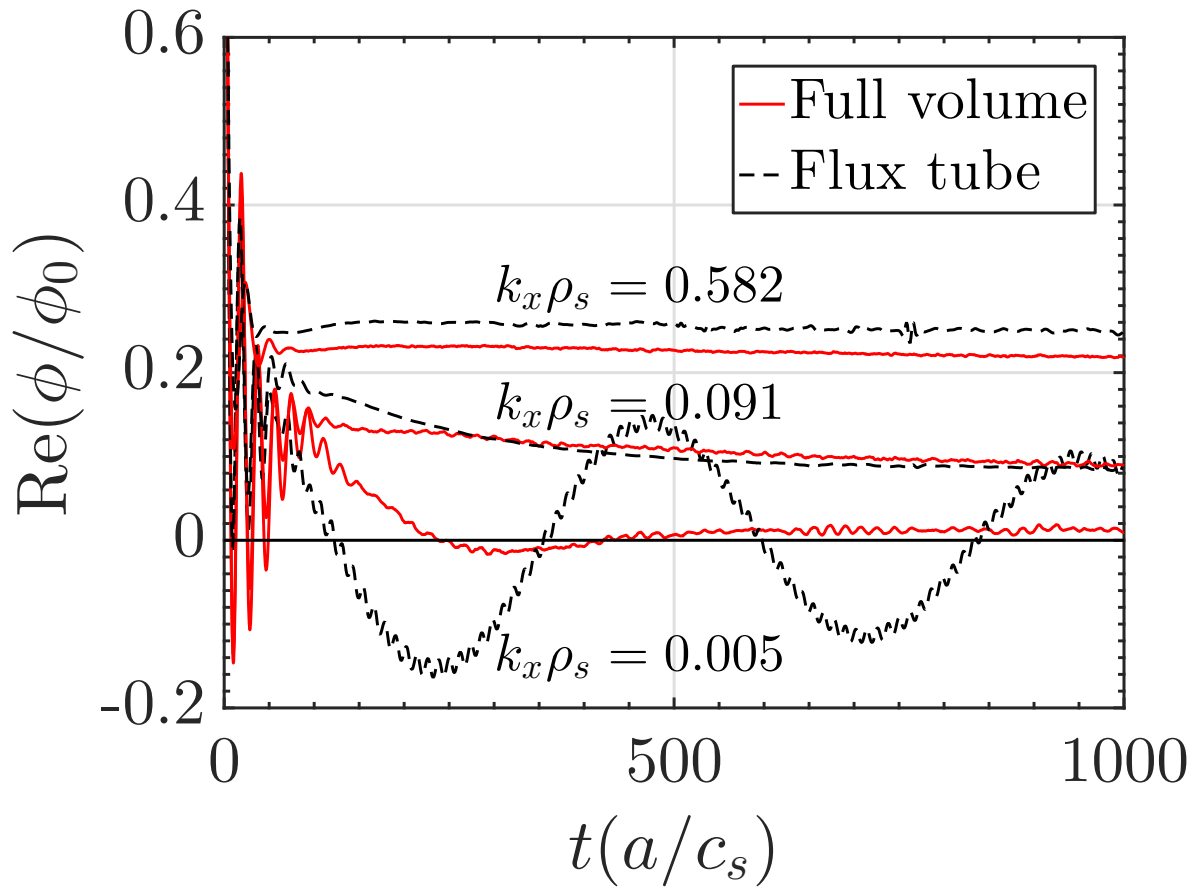}
  \caption{Comparison of zonal flow evolution in NCSX for full-volume and flux-tube calculations. Three different \kx/ demonstrate differences in the residual, long-time decay rate, and zonal flow oscillation damping. Calculations agree on the GAM frequency, and residuals match at $\krho/=0.091$. Flux-tube calculations here use $\npol/=4$.\label{fig:qa_global_t}}
\end{figure}

The short-time evolution of the zonal flow
is arguably more important than the long-time zonal flow residual
for turbulence saturation,
as turbulent correlation times are on the order of $\tau \sim 10 \t/$.
The time traces for several \kx/ are compared 
in Fig.~\ref{fig:qa_surface_t} for the flux-surface and flux-tube calculations,
and in Fig.~\ref{fig:qa_global_t} for the full-volume and flux-tube calculations.

\begin{figure}
  \includegraphics[width=\linewidth]{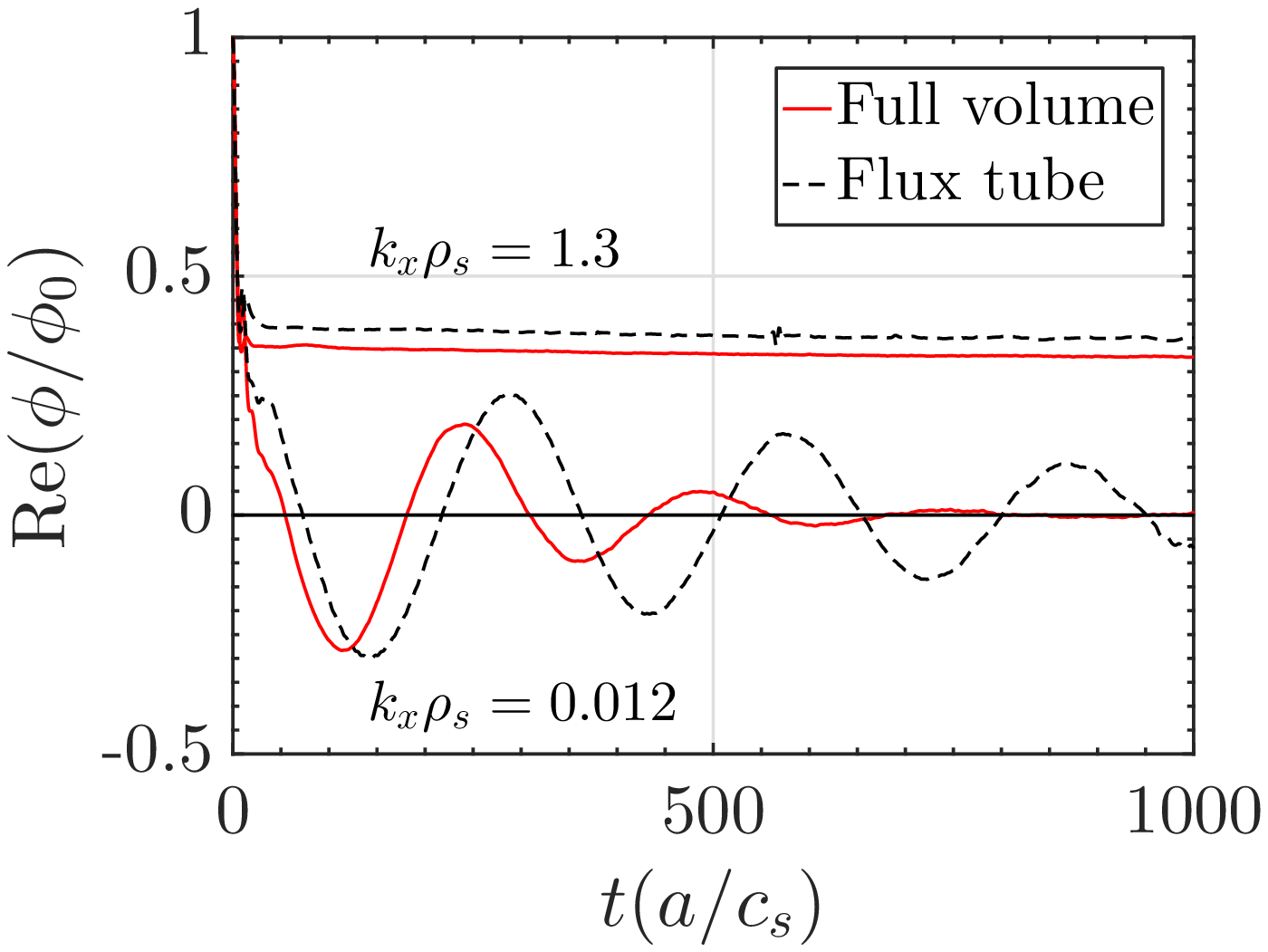}
  \caption{Comparison of zonal flow evolution in HSX Mirror configuration for full-volume and flux-tube calculations. Two different \kx/ demonstrate differences in the residual and the zonal flow oscillation frequency and damping, but similar time evolution at corresponding \kx/. Flux-tube calculations here use $\npol/=4$.\label{fig:f14_global_t}}
\end{figure}

Fig.~\ref{fig:qa_surface_t} shows that there is little difference
between flux-tube and flux-surface calculations. 
This only holds true for long enough flux tubes
as measured by convergence in \npol/,
as will be discussed in Sec.~\ref{sec:fluxtube}.
Evidently, the $k_y = 0$ mode in the flux tube
samples sufficient geometry through the parallel domain
that the same physics is retained as for the true flux-surface average.

In Fig.~\ref{fig:qa_global_t}, 
zonal flow oscillations can be identified for the smallest \kx/,
and there is significant long-time decay of the zonal flow for mid-\kx/.
The zonal flow decay in realistic NCSX geometry
is characteristic of that in a un-optimized stellarator.
The difference between flux-tube and full-volume calculations
is much more significant 
than that between flux-tube and flux-surface calculations.
At high $\krho/ \gtrsim 0.5$, 
the short-time decay due to the polarization drift reduces the zonal flow
to a smaller value in the full-volume calculation.
There is no difference in the long-time decay,
and so the zonal flow residual is smaller in the full-volume calculation at high \kx/. 
The zonal flow residual converges during the long-time evolution
for moderate $\krho/ \approx 0.1$,
but with slightly different decay properties in the two calculations.
Again, the zonal flow initially decays to a smaller amplitude
in the full-volume calculation, 
but the long-time decay is larger in the flux-tube calculations
such that the residual zonal flow is the same. 
The GAM frequency is consistent between full-volume and flux-tube calculations
but the amplitude is slightly smaller, 
or, alternatively, the GAM oscillation damping is slightly stronger
in the flux-tube calculation. 
\changed{In Fig.~\ref{fig:qa_global_t}, 
zonal flow oscillations are visible only for very small $\krho/ << 0.1$.}
\changed{Z}onal flow oscillations \changed{that are sustained} in the smallest-\kx/ \changed{flux-tube} calculations
are quickly damped out in the full-volume calculation.
We include flux-tube and full-volume calculations in the HSX Mirror configuration
in Fig.~\ref{fig:f14_global_t}, where
the zonal flow oscillations are not damped as quickly
and can be more easily compared. 
At higher \kx/, there again exists only a small displacement of the residual,
whereas on very large scales
the full-volume calculation shows significantly larger damping of the zonal flow oscillations,
but also a slight increase in the oscillation frequency. 

Overall, good agreement is observed between flux-tube and flux-surface calculations \changed{in the NCSX geometry.} 
Full-volume calculations differ at system-size scales 
where global effects become important
but otherwise show fair agreement with radially local frameworks.
\changed{This agreement only holds for sufficiently long flux tubes,
as is discussed in the next section.
A similar comparison of flux-tube and full-volume calculations
is discussed for HSX in Sec.~\ref{sec:fluxtube_qhs}.}

\section{Zonal flow response in different flux tubes}
\label{sec:fluxtube}

\begin{figure}
  \includegraphics[width=\linewidth]{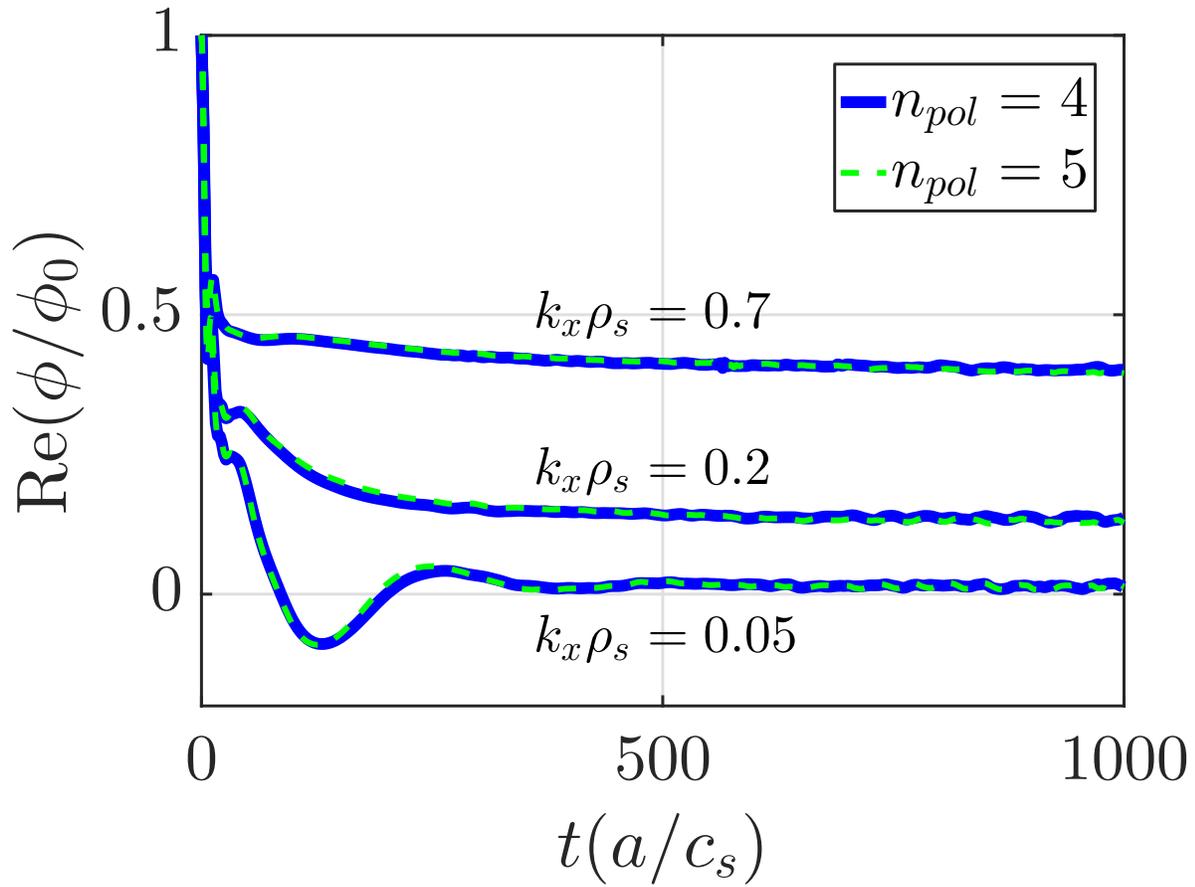}
  \caption{Zonal flow evolution for $k_x\rho_s=[0.05,0.2,0.7]$ with $\npol/=[4,5]$ in the HSX Mirror configuration. The value of $q \npol/ N$ is $14.96$ for $\npol/=4$ and $18.7$ for $\npol/=5$. Convergence is seen in \npol/ despite the non-integer value of $q \npol/ N$.\label{fig:npol_check}}
\end{figure}

\begin{figure}
  \includegraphics[width=\linewidth]{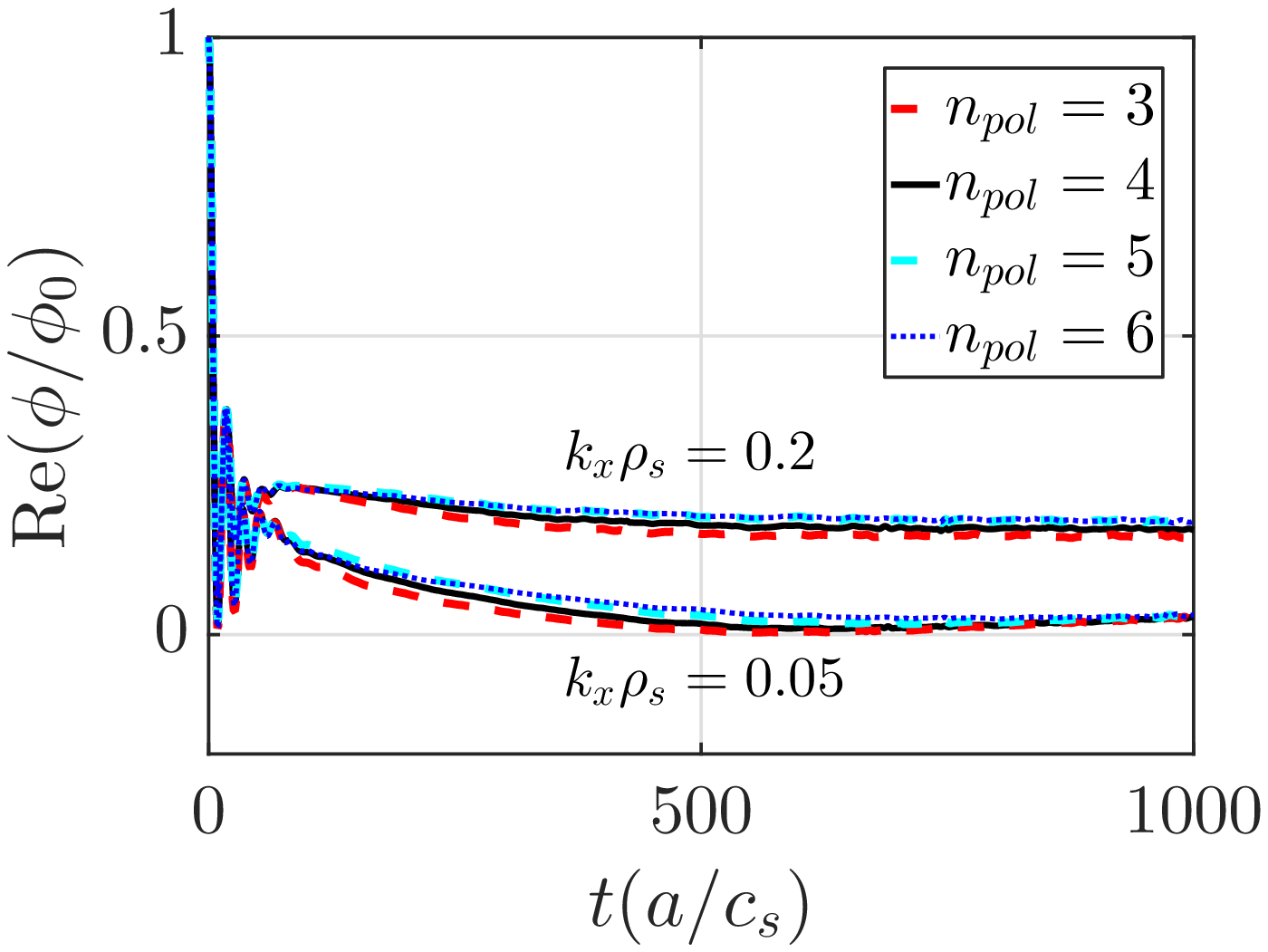}
  \caption{Zonal flow evolution from flux tubes at $s=0.54$ in NCSX, where $q=0.5714\approx 7/4$. The condition $q \npol/ N = 21$ for $\npol/ = 4$. However, convergence is achieved for $\npol/=[3,5,6]$, where $q \npol/ N = [15.75,26.25,31.5]$.\label{fig:qnN_check}}
\end{figure}

The calculation of the zonal flow response in a flux tube is 
computationally cheaper compared with flux-surface or full-volume calculations,
but is limited to the geometry information from a single field line.
As the zonal flow is toroidally and poloidally symmetric
and its dynamics depend on both 
bounce averages of the trapped particle radial drift
and flux-surface averages over the quasineutrality equation\cite{Monreal2016:PPCF-rzft},
a measurement of the zonal flow must be the same for any point on the flux surface.
In a \changetwo{general stellarator flux tube,}
each $\theta^* = [-\pi,\pi]$ flux tube is unique
and contains different geometry information\changetwo{.
True geometric periodicity requires that $q \npol/ N = \mathrm{integer}$, 
as discussed in Sec.~\ref{sec:gene_calcs}.
With $q=0.9413$ in QHS and $q=0.9349$ in Mirror,
the HSX flux tubes at $\npol/=4$ closely approach the integer condition
with $q \npol/ N = 15.06$ in the QHS configuration 
and $q \npol/ N = 14.96$ in the Mirror configuration. 
The $\npol/=8$ flux tube in NCSX is also close to an integer,
with $q \npol/ N = 42.93$.
For the HSX Mirror case shown in Fig.~\ref{fig:npol_check},
the same results are obtained within the usual convergence thresholds
for $\npol/=4$ and $\npol/=5$, where $q \npol/ N = 18.7$.
Similarly, the condition is matched much more closely in Fig.~\ref{fig:qnN_check}
by minimally changing the radial position for an NCSX flux tube to $s=0.54$,
such that $q = 0.5714 \approx 7/4$ and $q \npol/ N = 21$ for $\npol/ = 4$.
Calculations at $\npol/=3,5,6$ again converge to the $\npol/=4$ flux tube
despite the non-integer value of $q \npol/ N$. 
This is consistent with Ref.~\onlinecite{Martin2018:PPCF-pbct} which showed that 
zonal flow residuals converged for a long enough flux tube,
regardless of the boundary condition.}  
We compare calculations in different flux tubes on the same surface,
and extend those flux tubes
\changetwo{to see convergence on the surface and} capture all relevant zonal flow effects.
%
%
%

\subsection{Comparison of response in two flux tubes in QHS}
\label{sec:fluxtube_qhs}

\begin{figure}
  \includegraphics[width=\linewidth]{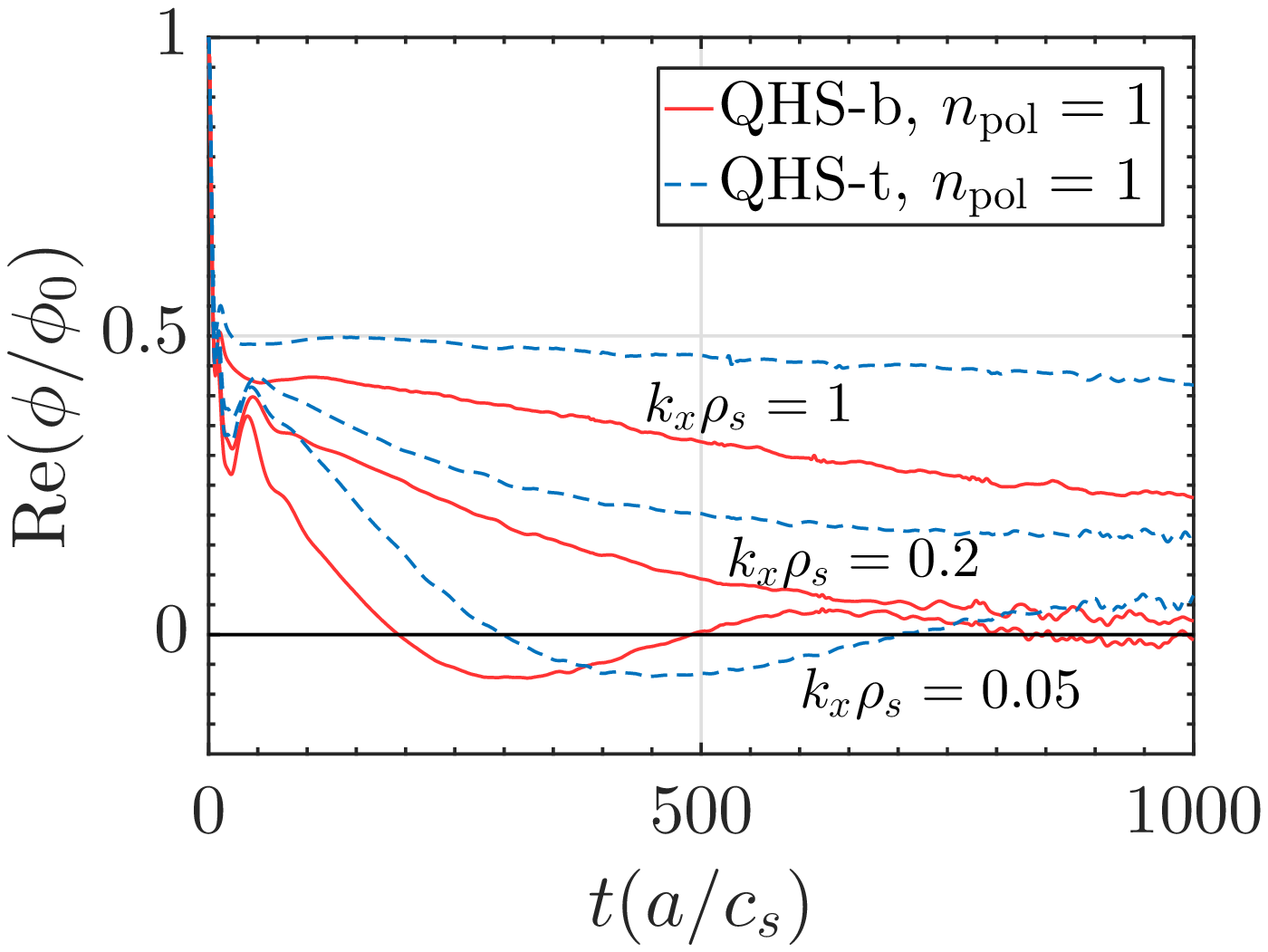}
  \caption{Zonal flow timetrace from two flux tubes with length of one poloidal turn. Decay times, zonal flow oscillations, and residual are different between flux tubes.\label{fig:cmp_qhs_tubes_npol1}}
\end{figure}

Unlike a tokamak, two flux tubes 
on the same surface in a stellarator
do not share the same geometry information. 
Here, we examine the zonal flow response
in two different flux tubes of the QHS configuration of HSX \changed{
introduced in Sec.~\ref{sec:geometry}.
The QHS-b ``bean'' flux tube is centered at $\alpha=0$,
while the QHS-t ``triangle'' flux tube is centered at $\alpha=\pi/4$.}
In Fig.~\ref{fig:cmp_qhs_tubes_npol1},
calculations of zonal flow damping with a flux-tube length of one poloidal turn
show large differences between the two flux tubes.
The zonal flow amplitude and decay rate are different,
and at small \kx/, the zonal flow oscillation frequency is larger in the QHS-b flux tube.
Neither individual $\npol/=1$ flux tube matches the zonal flow damping in a full-volume calculation.

\begin{figure}
  \includegraphics[width=\linewidth]{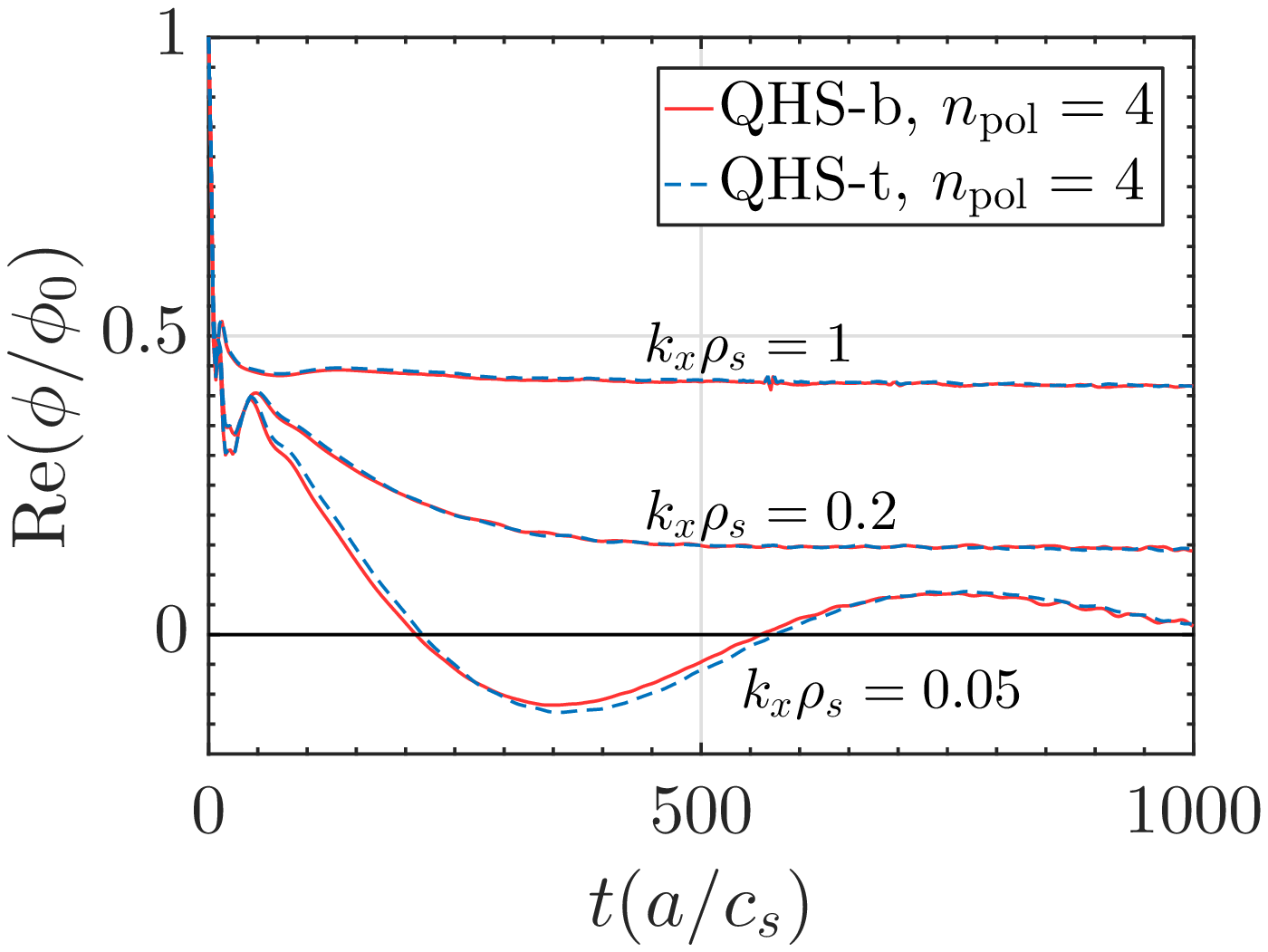}
  \caption{Zonal flow timetrace from two QHS flux tubes with length of four poloidal turns. The zonal flow response agrees in all fit parameters when flux tubes are extended.\label{fig:cmp_qhs_tubes_npol4}}
\end{figure}

\begin{figure}
  \includegraphics[width=\linewidth]{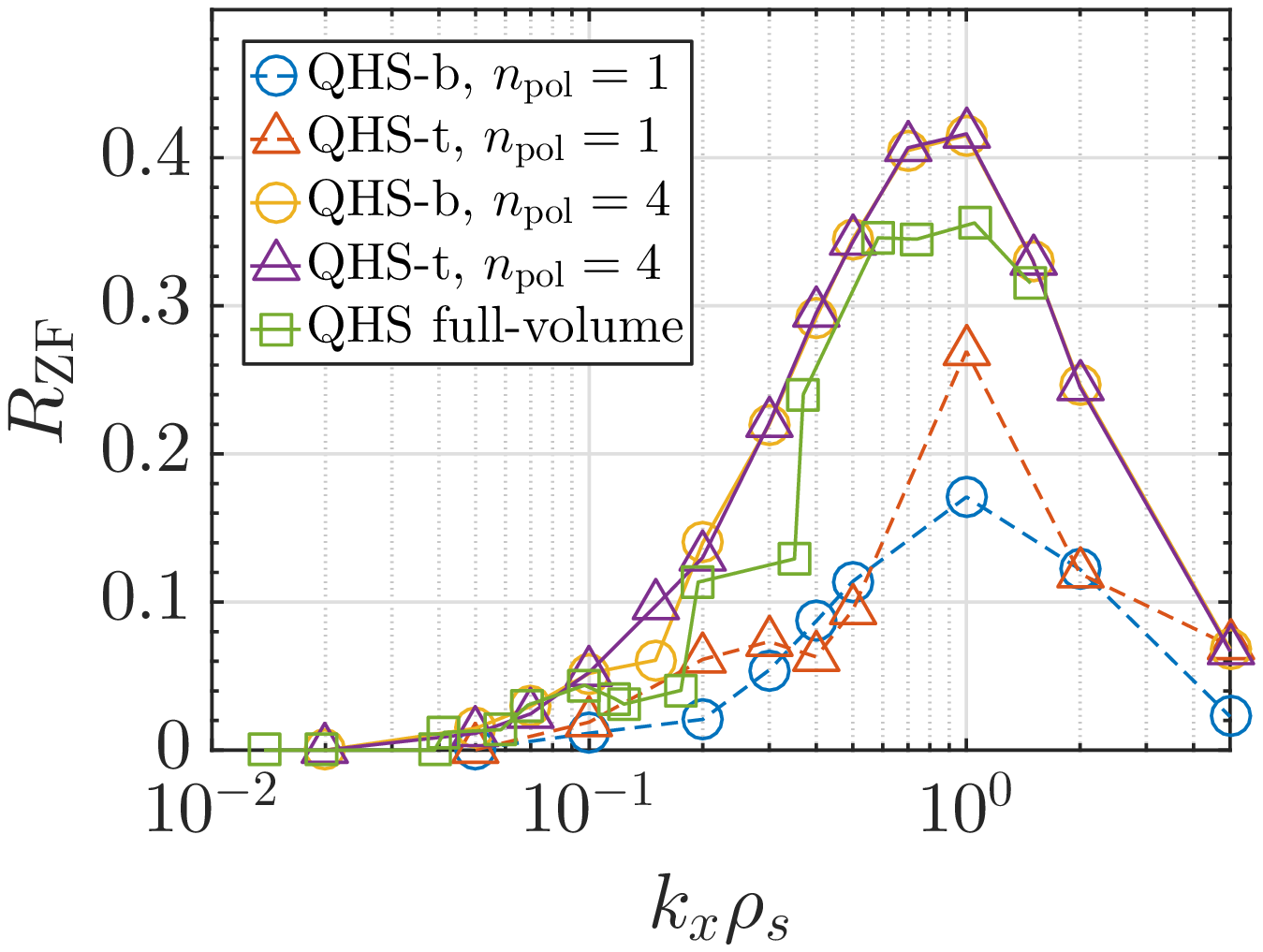}
  \caption{Zonal flow residuals for the QHS-b and QHS-t flux tubes. Shown are results for full-volume geometry and for flux tubes of two different lengths. Only the residuals from the extended flux tubes show agreement.\label{fig:cmp_qhs_tubes_Rzf}} 
\end{figure}

Geometry information can be added
by extending the flux tube to multiple poloidal turns along the field line.
The time traces from both flux tubes match
when the flux tube is extended to 4 poloidal turns in Fig.~\ref{fig:cmp_qhs_tubes_npol4}.
Furthermore, the same holds true for the zonal flow residual
across the \kx/ spectrum in Fig.~\ref{fig:cmp_qhs_tubes_Rzf}.
At $\npol/=1$, both the ``bean'' and the ``triangle'' flux tubes
demonstrate much more decay of the zonal flow residual
than the full-volume calculation.
As results from both flux tubes change as the flux tube is extended,
neither flux tube has enough information to calculate
the zonal flow damping correctly at one poloidal turn.
However, the flux tube recovers the flux-surface average
at four poloidal turns,
and both flux tubes produce the same zonal flow residual.
All other HSX flux-tube calculations in this paper
use the $\npol/=4$ ``bean'' flux tube.

\subsection{Comparison of response in three flux tubes in NCSX}
\label{sec:fluxtube_ncsx}

\begin{figure}
  \includegraphics[width=\linewidth]{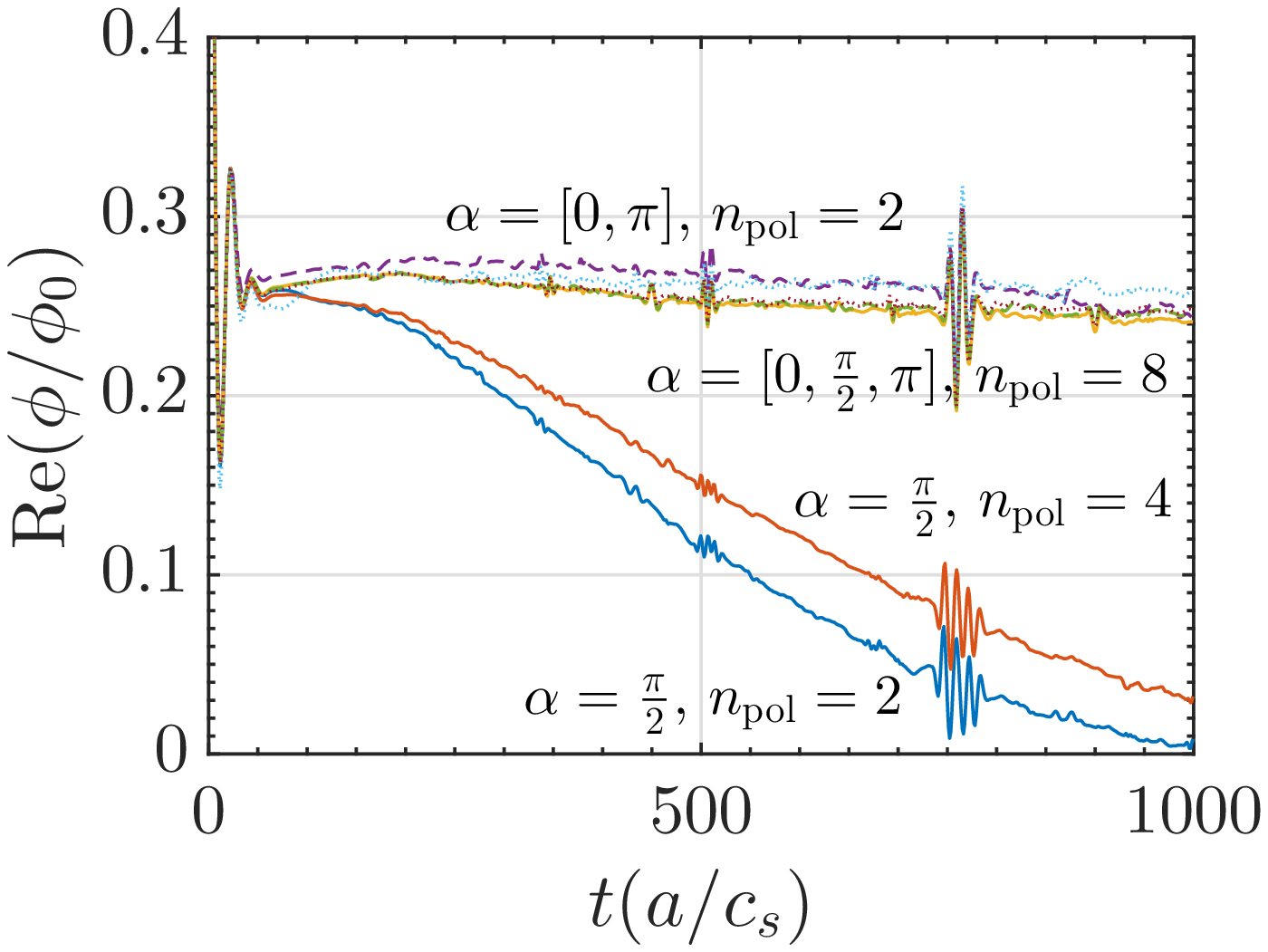}%
  \caption{Zonal flow damping in NCSX flux tubes for $\krho/=0.5$, where $\alpha=0$ is dashed, $\alpha=\pi/2$ is solid, and $\alpha=\pi$ is dotted. At $\npol/=2$, the $\alpha=[0,\pi]$ flux tubes agree, but the $\alpha=\pi/2$ flux tube decays to zero residual. All three flux tubes produce the same result at $\npol/=8$. The spike at $t=700$ is a numerical recurrence effect dependent on the velocity space resolution, and does not affect the interpretation.\label{fig:cmp_QA_convergence}}
\end{figure}

\begin{figure}
  \includegraphics[width=\linewidth]{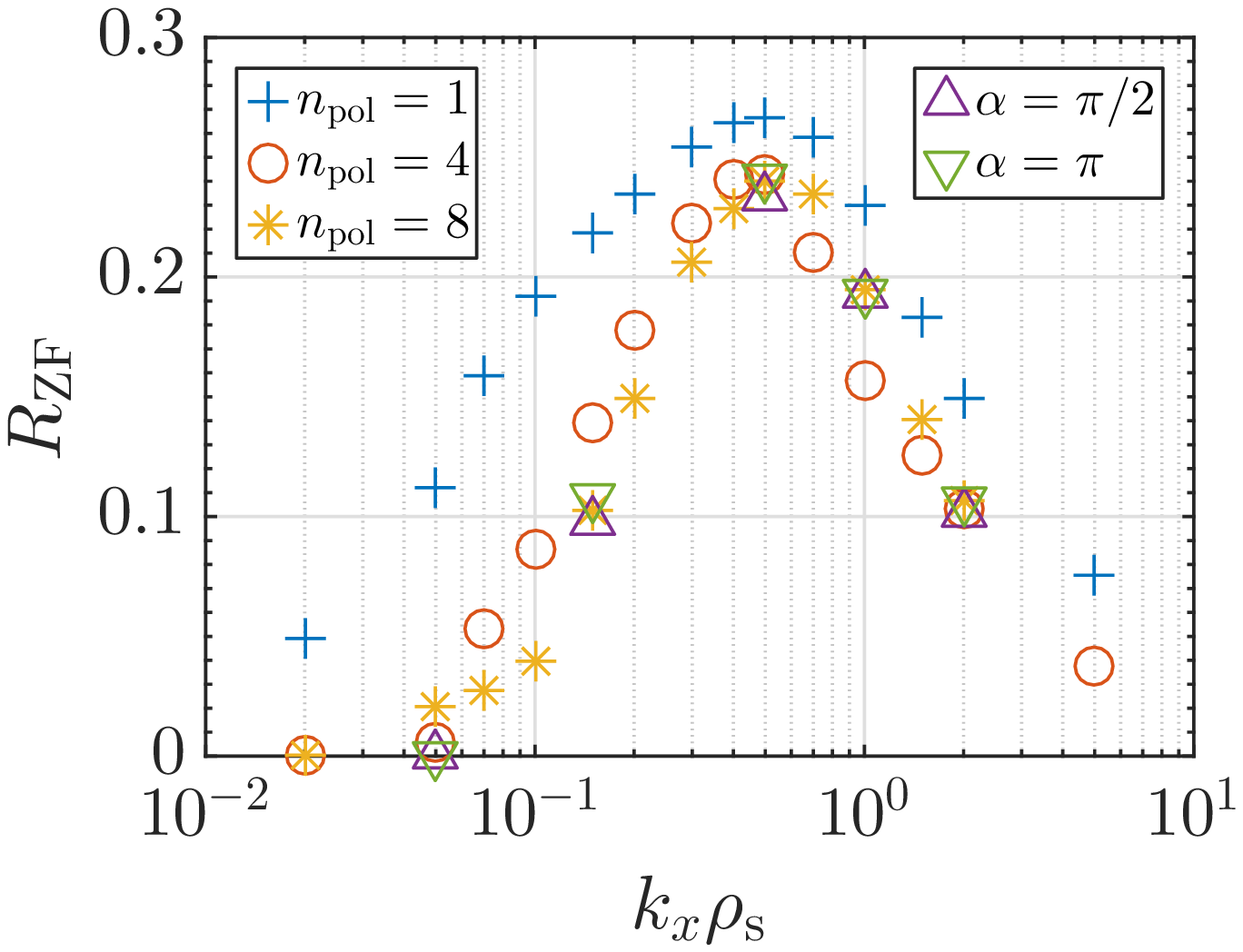}
  \caption{\kx/ spectra of the zonal flow residual $R_\mathrm{ZF}$ in the $\alpha=0$ flux tube of NCSX. At low \kx/, $R_\mathrm{ZF}$ depends strongly on \npol/. Points for the $\alpha = \pi/2, \pi$ flux tubes are plotted for $\npol/=8$, where all flux tubes converge to the same $R_\mathrm{ZF}$. \label{fig:cmp_QA_Rzf}}
\end{figure}

Three flux tubes are examined in the NCSX configuration.
The $\alpha=0$ and $\alpha=\pi$ flux tubes are 
symmetric about the midpoint $z=0$,
while the $\alpha=\pi/2$ flux tube is not.
As seen in Fig.~\ref{fig:cmp_QA_convergence},
the zonal flow damping is very different
in the $\alpha=\pi/2$ flux tube.
No zonal flow residual is supported
when the flux tube is fewer than eight poloidal turns long. 
The symmetric flux tubes
capture the zonal flow damping
at just two poloidal turns.
The poloidal distance between turns
is larger in NCSX than in HSX 
due to the difference in rotational transform.
This larger poloidal step size
samples broad variation on the flux surface,
but could under-sample geometry variations
that are smaller scale than the poloidal space between turns.
Note that with a rotational transform of about one half of that in HSX,
the toroidal length of a two-poloidal-turn flux tube in NCSX is
roughly the same as a four-poloidal-turn flux tube in HSX.
However, convergence of the zonal flow residual for low $\krho/<0.2$
imposes an even more restrictive requirement on flux-tube length,
as seen in Fig.~\ref{fig:cmp_QA_Rzf}.
Only at eight poloidal turns do all flux tubes produce the same zonal flow residual for all \kx/.

\section{Comparison of configurations: QHS, Mirror, and NCSX}
\label{sec:global}

\begin{figure}
  \centering
  \includegraphics[width=\linewidth]{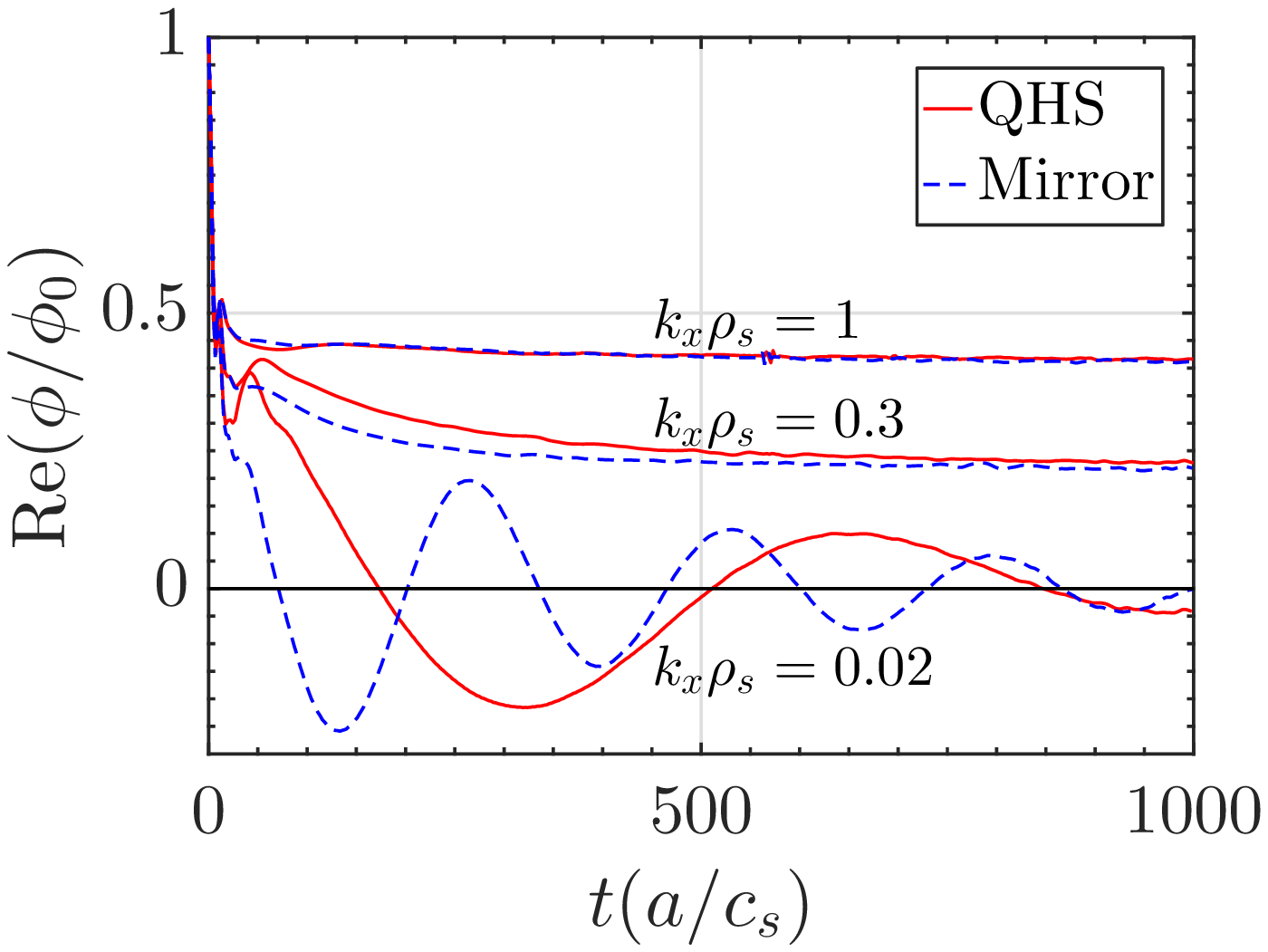}
  \caption{The time evolution of the zonal flow in the two HSX configurations, with $\npol/=4$ flux tubes. The Mirror configuration has higher frequency zonal flow oscillations and faster long-time damping, as expected based on its reduced quasi-symmetry. \label{fig:zonal-t}}
\end{figure}

The QHS and Mirror configurations of HSX have been designed
specifically to study differences in neoclassical transport and flow damping. 
As discussed in Sec.~\ref{sec:theory},
the zonal flow long-time damping and oscillation frequency
are related to neoclassical transport.
According to theory\cite{Sugama2006:PoP-cdzf,Helander2011:PPCF-ozfs},
the more optimized QHS configuration
should exhibit lower-frequency zonal flow oscillations
as well as slower long-time decay to the residual level.
These expectations are verified in Fig.~\ref{fig:zonal-t}.
The zonal flow oscillation frequency is higher by a factor of 2.5 in Mirror than QHS at $\krho/=0.02$,
and the long-time damping is significantly faster at $\krho/=0.3$. 

\begin{figure}
  \includegraphics[width=\linewidth]{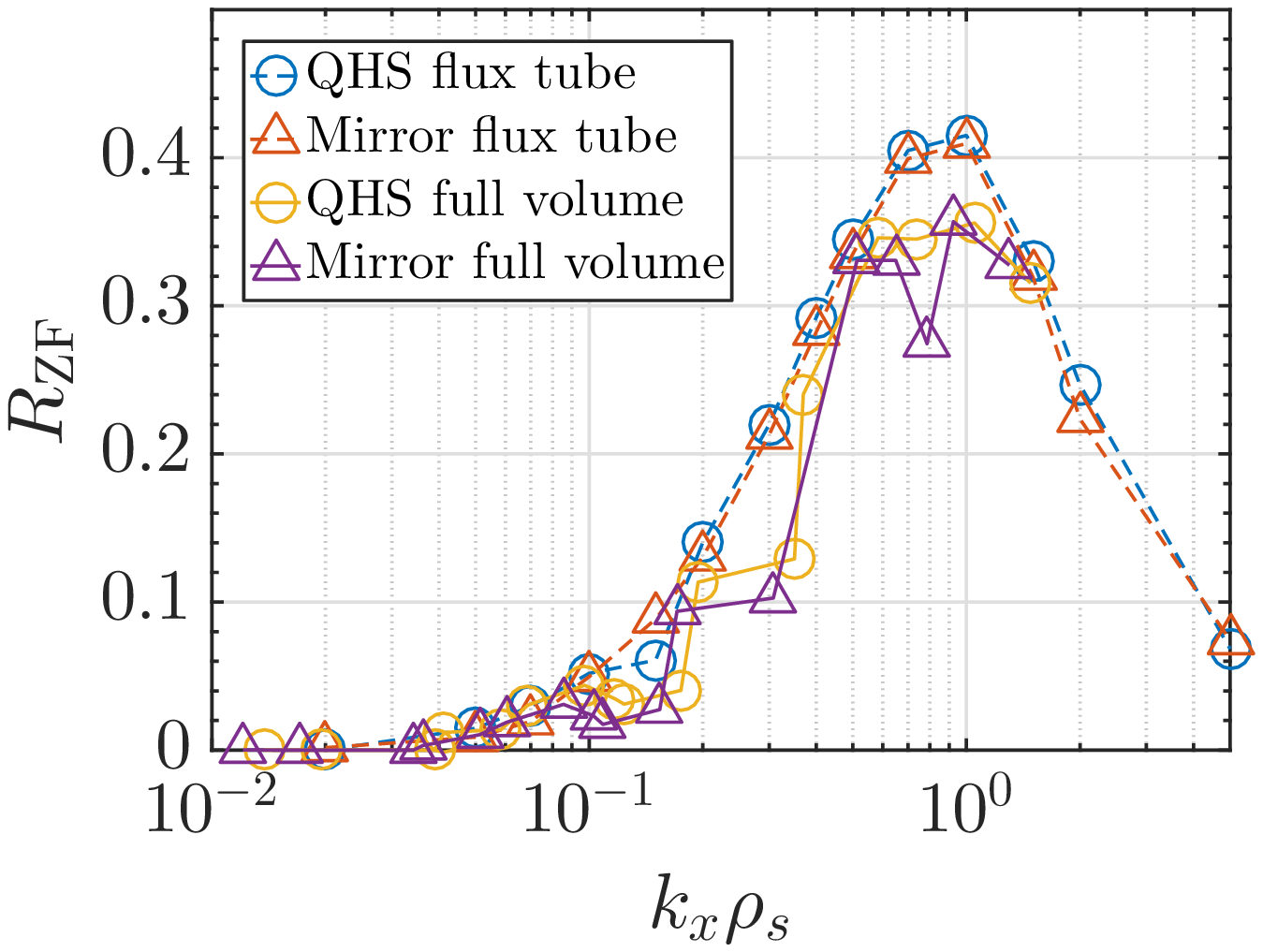}
  \caption{Zonal flow residual \kx/ spectra for QHS and Mirror. The flux tubes for \gene/ calculations are 4 poloidal turns long, and show good agreement with full-volume calculations. In both flux-tube and full-volume calculations, there is no significant difference between QHS and Mirror. \label{fig:cmp_qhs_f14}}
\end{figure}

As observed in Fig.~\ref{fig:cmp_qhs_f14},
the zonal flow residual does not differ between QHS and Mirror.
The Rosenbluth-Hinton residual
in Eq.~(\ref{eq:RH_residual})
depends primarily on the safety factor $q$,
a consequence of the ratio of the banana-induced polarization
relative to the gyro-orbit-induced polarization\cite{Sugama2006:PoP-cdzf}.
In a non-axisymmetric system, 
additional trapped particles further modify the zonal flow evolution
through their polarization and radial drift.
While the radial drift is important for the time evolution,
the zonal flow residual primarily depends upon the polarization effects provided by the trapped particles\cite{Xanthopoulos2011:PRL-zfdc}. 
The broken symmetry of the mirror configuration increases the trapped particle radial drift,
as demonstrated by the zonal flow oscillation frequency and damping,
but does not change the zonal flow residual. 
We conclude that the helically trapped particles
dominate the polarization drift and set the zonal flow residual in both systems.

As compared to NCSX,
the QHS configuration produces less trapped-particle radial drift
and has a lower zonal flow oscillation frequency,
while the  Mirror configuration produces more
and has a higher oscillation frequency. 
GAMs are damped more slowly in the NCSX configurations,
due to the larger safety factor $q$,
and GAM oscillations can be seen in Figs.~\ref{fig:qa_surface_t} and \ref{fig:qa_global_t}
but are barely identifiable in any HSX timetraces. 
\changed{In comparing the zonal flow residual in Figs.~\ref{fig:cmp_QA_Rzf} and \ref{fig:cmp_qhs_f14},
t}he peak residual is smaller in the NCSX configuration,
but the peak location is found at a different \kx/.
The residual in HSX configurations peaks at $\krho/\approx 1$,
similar to Wendelstein 7-X,
while the NCSX configuration peaks at $\krho/\approx 0.5$,
similar to the tokamak in Ref.~\cite{Monreal2016:PPCF-rzft}.

\begin{figure}
  \includegraphics[width=\linewidth]{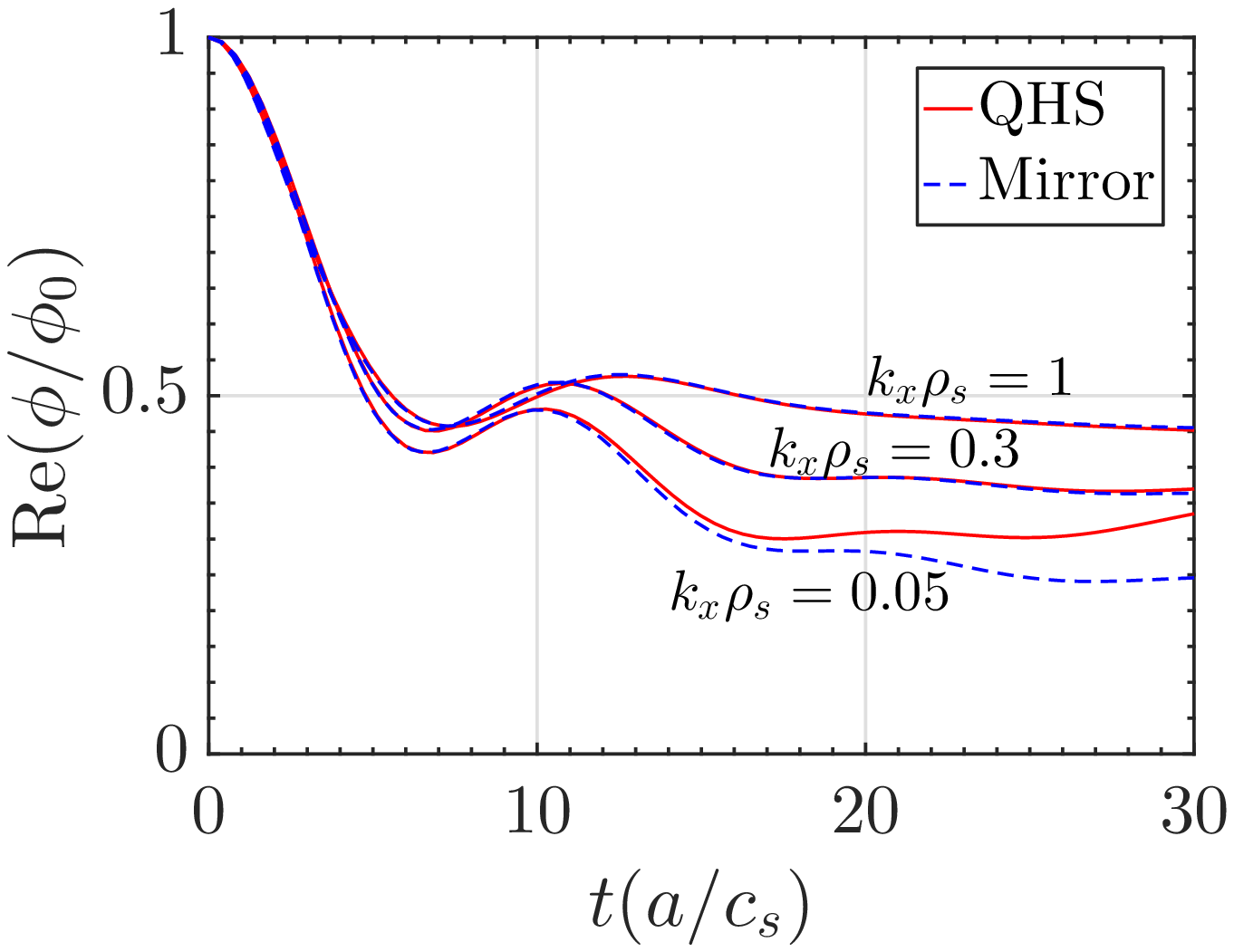}
  \caption{The time evolution of the zonal flow for very short times, comparable to the turbulent correlation time. Effectively, there is no difference between QHS and Mirror except at the smallest \kx/, and only after about 20 \t/. \label{fig:cmp_qhs_f14_short}}
\end{figure}

Calculation of the zonal flow decay in HSX
captures the expected neoclassical effects on decay rate and oscillation frequency.
However, the saturation of drift wave turbulence 
is a strong motivation for the study of zonal flows. 
Turbulence will transfer energy and reorganize the system
within a correlation time,
effectively resetting the zonal flow time evolution.
In nonlinear simulations of trapped electron mode (TEM) turbulence in HSX\cite{Smoniewski:GFpaper},
the correlation time is on the order of 10 \t/. 
In Fig.~\ref{fig:cmp_qhs_f14_short},
the short-time damping of the zonal flow is plotted,
but again, there is no difference between the QHS and Mirror configurations.
Depending on driving gradients,
the heat flux from nonlinear TEM turbulence simulations 
differs between these configurations, 
and is not explained by the linear growth of the most unstable mode\cite{Smoniewski:GFpaper}.
If a difference in heat fluxes between configurations 
is due to the linear collisionless zonal flow dynamics,
it is not a simple relation to either the very short-time dynamics or the long-time residual.
Instead, it could be hypothesized that, if such a relation exists, 
it would stem from a shift in characteristic \kx/ of linear drive physics, 
which would affect which zonal flow acts to saturate the turbulence.
We do not address the question of the effect of an external radial electric field on the zonal flow,
and a difference of the ambipolar radial electric field between QHS and Mirror 
could lead to important differences in the zonal flow decay.

Having demonstrated that simulations confirm 
the link between broken symmetry and a faster erosion of the zonal flow residual,
a link can be established to a similar effect in axisymmetric systems. 
There, resonant magnetic perturbations, 
whether created by external coils\cite{Williams2020:NF-irmp},
by magneto-hydrodynamic activity\cite{Williams2017:PoP-ttzf},
or by microturbulence itself\cite{Pueschel2013:PRL-ehfg},
erode the zonal flow residual
and lead to increased turbulent transport\cite{Terry2013:PoP-emfr,Pueschel2013:PoP-phmn}. 
However, erosion time scales in these scenarios
were on the order of the turbulent correlation time,
giving further credence to the idea that the long-time decay
present in the systems investigated here
is unlikely to affect transport directly.

\section{Summary}

We have presented calculations of
linear zonal flow damping
in  quasi-symmetric stellarators.
In the geometries of NCSX and HSX,
the time evolution is dictated by
the typical characteristics of non-axisymmetric devices.
The zonal flow residual vanishes
for small \kx/,
the zonal flow undergoes long-time decay to the residual,
and zonal flow oscillations occur.
Calculations are performed
in full-volume, flux-surface, and flux-tube geometries.
A sufficiently long flux tube reproduces the full-volume residual
and flux-surface time-dependence,
suggesting that parallel dynamics in an appropriate flux tube
can approximate the flux-surface average.
While $\npol/=4$ and $\npol/=8$ is sufficient to recover flux-surface results in these two configurations,
the required flux-tube length is configuration-dependent and cannot be taken as a general rule. 
It should be noted that both flux-tube and flux-surface calculations
exhibit slightly less decay during the short-time polarization drift
than a full-volume calculation.
On the other hand,
the damping of the zonal flow oscillation is greater in the full-volume calculation.
The zonal flow oscillation is only visible at small \kx/,
where the full-volume calculation supports a larger residual
than local representations. 
This is likely due to the breakdown of the radially-local approximation
as \kx/ approaches the system size. 

The collisionless zonal flow decay examined here
cannot be correlated to the nonlinear turbulent transport
without further information.
Nonlinear simulations of TEM in the QHS and Mirror configurations 
produce different heat fluxes\cite{Smoniewski:GFpaper},
but the zonal flow residual at finite \kx/ 
shows no difference between QHS and Mirror. 
Given the short timescale of a turbulent correlation time,
the short-time damping of the zonal flow may be more
relevant to the saturation of turbulence. 
The polarization drift dominates 
the short-time zonal flow damping,
and there is no difference in the time evolution
of the QHS and Mirror configurations
until the zonal flow oscillation becomes significant.
The HSX QHS and Mirror configurations clearly demonstrate
a difference in zonal flow oscillations 
and long-time decay,
but these differences follow the expected dependence 
on the neoclassical radial drift.
Configurations with a larger radial drift
have a higher oscillation frequency and slower long-time decay.
These quantities cannot be related to the full zonal flow evolution
without also directly relating to the neoclassical optimization.
In addition, any extrapolation from linear zonal flow damping to nonlinear heat flux
requires an understanding of which \kx/ are important
for energy transfer in the specific system under study.

Future work should include external radial electric fields,
which can strongly modify the zonal flow decay and residual. 
The radial electric field in a stellarator is usually determined
by an ambipolarity constraint on neoclassical transport,
which can differ between configurations
but requires knowledge of density and temperature profiles.

\begin{acknowledgments}
The authors would like to acknowledge helpful discussions with C.C.~Hegna. 
This research has been supported by US DOE Grants DE-FG02-93ER54222 and DE-FG02-04ER54742
and used resources of the National Energy Research Scientific Computing Center (NERSC),
a U.S. Department of Energy Office of Science User Facility
operated under Contract No. DE-AC02-05CH11231.
This work has been partially funded by the Ministerio de Econom\'ia y Competitividad of Spain 
under projects ENE2015-70142-P and PGC2018-095307-B-I00.
The authors acknowledge the computer resources at Mare Nostrum IV
and the technical support
provided by the Barcelona Supercomputing Center
and the CIEMAT computing center.
This work has been carried out within the framework of the EUROfusion Consortium 
and has received funding from the Euratom research and training programme
2014-2018 and 2019-2020 under grant agreement No 633053.
The views and opinions expressed herein do not necessarily reflect those of the European Commission.
\end{acknowledgments}

\section*{DATA AVAILABILITY}
The data supporting the findings of this study is available from the authors upon reasonable request.

\end{document}